\PassOptionsToPackage{unicode}{hyperref}
\PassOptionsToPackage{hyphens}{url}
\PassOptionsToPackage{dvipsnames,svgnames,x11names}{xcolor}
\documentclass[
  12pt]{article}

\usepackage{float}
\usepackage{amsfonts}
\usepackage{mathrsfs}
\usepackage{hyperref}
\usepackage{xr}

\usepackage{amsbsy}
\usepackage{amsmath}
\usepackage{amssymb}
\usepackage{multirow}
\usepackage{graphicx}
\usepackage{varwidth}
\usepackage{bm}
\usepackage{color}
\usepackage{hyperref}
\usepackage{rotating}
\usepackage{graphicx}
\usepackage{epsfig}
\usepackage{epstopdf}
\usepackage{appendix}
\usepackage{ulem}
\usepackage{booktabs}       
\usepackage{amsfonts}       
\usepackage{nicefrac}       
\usepackage{microtype}      
\usepackage{xcolor}         
\usepackage{enumitem}
\usepackage[ruled, vlined, linesnumbered]{algorithm2e}
\SetAlCapFnt{\footnotesize}      
\usepackage{booktabs}
\usepackage{float} 
\usepackage{subfigure}
\usepackage{overpic}
\usepackage[flushleft]{threeparttable}
\usepackage[english]{babel}
\usepackage{amsthm}
\usepackage{indentfirst}
\usepackage{wrapfig}

\newtheorem{thm}{Theorem}

\newtheorem{corollary}{Corollary}

\newtheorem{dfn}{Definition}
\newtheorem{con}{Condition}

\newcommand{\btheta}{\boldsymbol{\theta}}

\newcommand{\bvmu}{\boldsymbol{\mu}}

\newcommand{\balpha}{\boldsymbol{\alpha}}

\newcommand{\blambda}{\boldsymbol{\lambda}}
\newcommand{\brho}{\boldsymbol{\rho}}

\newcommand{\bSigma}{\boldsymbol{\Sigma}}

\newcommand{\bs}{\boldsymbol{s}}

\newcommand{\by}{\boldsymbol{y}}
\newcommand{\bx}{\boldsymbol{x}}

\newcommand{\bz}{\boldsymbol{z}}

\newcommand{\ba}{\boldsymbol{a}}

\newcommand{\bu}{\boldsymbol{u}}
\newcommand{\bB}{\boldsymbol{B}}

\newcommand{\bW}{\boldsymbol{W}}
\newcommand{\bX}{\boldsymbol{X}}
\newcommand{\bU}{\boldsymbol{U}}
\newcommand{\bS}{\boldsymbol{S}}

\newcommand{\bV}{\boldsymbol{V}}

\newcommand{\bI}{\boldsymbol{I}}
\newcommand{\bJ}{\boldsymbol{J}}

\newcommand{\mcA}{\mathcal{A}}

\newcommand{\mcM}{\mathcal{M}}

\newcommand{\mcR}{\mathcal{R}}
\newcommand{\mcS}{\mathcal{S}}

\newcommand{\mcG}{\mathcal{G}}

\newcommand{\mcD}{\mathcal{D}}

\newcommand{\mcH}{\mathcal{H}}

\newcommand{\mcT}{\mathcal{T}}

\newcommand{\mbE}{\mathbb{E}}
\newcommand{\mbR}{\mathbb{R}}

\newcommand{\mcI}{\mathcal{I}}

\def\blue{\color{blue}}

\usepackage{iftex}
\ifPDFTeX
  \usepackage[T1]{fontenc}
  \usepackage[utf8]{inputenc}
  \usepackage{textcomp} 
\else 
  \usepackage{unicode-math}
  \defaultfontfeatures{Scale=MatchLowercase}
  \defaultfontfeatures[\rmfamily]{Ligatures=TeX,Scale=1}
\fi
\usepackage{lmodern}
\ifPDFTeX\else  
\fi
\IfFileExists{upquote.sty}{\usepackage{upquote}}{}
\IfFileExists{microtype.sty}{
  \usepackage[]{microtype}
  \UseMicrotypeSet[protrusion]{basicmath} 
}{}
\makeatletter
\@ifundefined{KOMAClassName}{
  \IfFileExists{parskip.sty}{%
    \usepackage{parskip}
  }{
    \setlength{\parindent}{0pt}
    \setlength{\parskip}{6pt plus 2pt minus 1pt}}
}{
  \KOMAoptions{parskip=half}}
\makeatother
\usepackage{xcolor}
\makeatletter
\ifx\paragraph\undefined\else
  \let\oldparagraph\paragraph
  \renewcommand{\paragraph}{
    \@ifstar
      \xxxParagraphStar
      \xxxParagraphNoStar
  }
  \newcommand{\xxxParagraphStar}[1]{\oldparagraph*{#1}\mbox{}}
  \newcommand{\xxxParagraphNoStar}[1]{\oldparagraph{#1}\mbox{}}
\fi
\ifx\subparagraph\undefined\else
  \let\oldsubparagraph\subparagraph
  \renewcommand{\subparagraph}{
    \@ifstar
      \xxxSubParagraphStar
      \xxxSubParagraphNoStar
  }
  \newcommand{\xxxSubParagraphStar}[1]{\oldsubparagraph*{#1}\mbox{}}
  \newcommand{\xxxSubParagraphNoStar}[1]{\oldsubparagraph{#1}\mbox{}}
\fi
\makeatother

\usepackage{longtable,booktabs,array}
\usepackage{calc} 
\usepackage{etoolbox}
\makeatletter
\patchcmd\longtable{\par}{\if@noskipsec\mbox{}\fi\par}{}{}
\makeatother
\IfFileExists{footnotehyper.sty}{\usepackage{footnotehyper}}{\usepackage{footnote}}
\makesavenoteenv{longtable}
\usepackage{graphicx}
\makeatletter
\def\maxwidth{\ifdim\Gin@nat@width>\linewidth\linewidth\else\Gin@nat@width\fi}
\def\maxheight{\ifdim\Gin@nat@height>\textheight\textheight\else\Gin@nat@height\fi}
\makeatother
\setkeys{Gin}{width=\maxwidth,height=\maxheight,keepaspectratio}
\makeatletter
\def\fps@figure{htbp}
\makeatother

\makeatletter
\@ifpackageloaded{caption}{}{\usepackage{caption}}
\AtBeginDocument{%
\ifdefined\contentsname
  \renewcommand*\contentsname{Table of contents}
\else
  \newcommand\contentsname{Table of contents}
\fi
\ifdefined\listfigurename
  \renewcommand*\listfigurename{List of Figures}
\else
  \newcommand\listfigurename{List of Figures}
\fi
\ifdefined\listtablename
  \renewcommand*\listtablename{List of Tables}
\else
  \newcommand\listtablename{List of Tables}
\fi
\ifdefined\figurename
  \renewcommand*\figurename{Figure}
\else
  \newcommand\figurename{Figure}
\fi
\ifdefined\tablename
  \renewcommand*\tablename{Table}
\else
  \newcommand\tablename{Table}
\fi
}
\@ifpackageloaded{float}{}{\usepackage{float}}
\floatstyle{ruled}
\@ifundefined{c@chapter}{\newfloat{codelisting}{h}{lop}}{\newfloat{codelisting}{h}{lop}[chapter]}
\floatname{codelisting}{Listing}

\makeatother
\makeatletter
\@ifpackageloaded{caption}{}{\usepackage{caption}}
\@ifpackageloaded{subcaption}{}{\usepackage{subcaption}}
\makeatother

\ifLuaTeX
  \usepackage{selnolig}  
\fi
\usepackage[]{natbib}
\usepackage{bookmark}

\IfFileExists{xurl.sty}{\usepackage{xurl}}{} 
\hypersetup{
  pdftitle={Title},
  pdfauthor={Author 1; Author 2},
  pdfkeywords={3 to 6 keywords, that do not appear in the title},
  colorlinks=true,
  linkcolor={blue},
  filecolor={Maroon},
  citecolor={Blue},
  urlcolor={Blue},
  pdfcreator={LaTeX via pandoc}}

\makeatletter
\newcommand*{\addFileDependency}[1]{
\typeout{(#1)}
\@addtofilelist{#1}
\IfFileExists{#1}{}{\typeout{No file #1.}}
}\makeatother

\newcommand*{\myexternaldocument}[1]{%
\externaldocument{#1}%
\addFileDependency{#1.tex}%
\addFileDependency{#1.aux}%
}

\myexternaldocument{DeepIn_SM20260306}

\newcommand{\anon}{1}

\begin{document}

\def\spacingset#1{\renewcommand{\baselinestretch}%
{#1}\small\normalsize} \spacingset{1}


\if1\anon
{
  \title{\bf Minimal Sufficient Representations \\ for Self-interpretable Deep Neural Networks
  \thanks{The research was supported by National Key R\&D Program of China (No.2022YFA1003702), National Natural Science Foundation of China (Nos.12426309, 12171374, 12371275), Sichuan Science and Technology Program, China (Grant No. 2025JDDJ0007), Opening Project Fund of National Facility for Translational Medicine (Shanghai), New Cornerstone Science Foundation, and Guanghua Talent Project of SWUFE.}
  }
  \date{}
  \author{Zhiyao Tan$^{1\dagger}$, Li Liu$^3$\thanks{Co-first authors.} \  and  Huazhen Lin$^{1,2}$\thanks{Corresponding author. Email address: \emph{linhz@swufe.edu.cn}.}  \vspace{.5cm}\\
      $^1$ Center of Statistical Research,  School of Statistics, and \\
      $^2$ New Cornerstone Science Laboratory,  \\
      Southwestern University of Finance and Economics, Chengdu, China\\
     $^3$ School of Mathematics and Statistics, Wuhan University, Wuhan, China}
  \maketitle
} \fi

\if0\anon
{
  \bigskip
  \bigskip
  \bigskip
  \begin{center}
    {\LARGE\bf Title}
\end{center}
  \medskip
} \fi

\bigskip
\begin{abstract}
Deep neural networks (DNNs) achieve remarkable predictive performance but remain difficult to interpret, largely due to  overparameterization that obscures the minimal structure required for interpretation.
Here we introduce DeepIn, a self-interpretable neural network framework that adaptively identifies and learns the minimal representation necessary for  preserving the full expressive capacity of standard DNNs.
We  show that DeepIn can correctly identify the minimal representation dimension, select relevant variables, and recover the minimal sufficient network architecture for prediction.  The resulting estimator achieves optimal non-asymptotic error rates that adapt to the learned minimal dimension, demonstrating that recovering minimal sufficient structure fundamentally improves generalization error. Building on these guarantees, we further develop hypothesis testing procedures for both selected variables and learned representations, bridging deep representation learning with formal statistical inference.
Across biomedical and vision benchmarks, DeepIn improves both predictive accuracy and interpretability, reducing error by up to 30\% on real-world datasets while automatically uncovering human-interpretable discriminative patterns.
Our results suggest that interpretability and statistical rigor can be embedded directly into deep architectures without sacrificing performance.
\end{abstract}

\noindent%
{\it Keywords:} Interpretable DNNs, High-dimensional learning, Representation learning, Variable selection, Hypothesis testing, Adaptive network architecture
\vfill

\newpage
\spacingset{1.8} 

\section{Introduction}

Modern deep neural networks (DNNs) are typically heavily overparameterized, containing far more degrees of freedom than are strictly necessary for prediction. Such redundancy can lead to unstable estimation, inflated variance,  substantial computational inefficiency, and heavy reliance on enormous amounts of training data, while obscuring the minimal structural components truly responsible for predictive performance. As a result, deep models often behave as black boxes: the intrinsic representation dimension underlying their decisions remains unknown, irrelevant features may be entangled with informative ones, and principled uncertainty quantification is rarely available. These limitations pose substantial challenges for deploying deep learning systems in high-stakes scientific and societal applications, for example,  medical diagnosis, financial risk assessment, and public policy evaluation, where transparency, interpretability, and statistical reliability are essential. 
A fundamental questions therefore arises:
Can we  recover the minimal sufficient  representations in DNNs  so that interpretability and statistical rigor can be incorporated into deep architectures without degrading predictive accuracy?

Recently, there has been growing interest in  interpreting DNNs. A common strategy relies on post-hoc interpretability methods \citep{ribeiro2016should, lundberg2017unified}, which employ surrogate interpretable models, such as linear models or decision trees, to locally approximate the behavior of trained DNNs. However,  these methods often produce inconsistent or even misleading explanations, as they capture only local approximations rather than the model's true internal mechanisms \citep{ghorbani2019interpretation}. Moreover, post-hoc explanations are detached from the model training process, implying that interpretability
cannot guide the model toward learning more transparent representations  \citep{rudin2019stop}.

To address the limitations of post-hoc interpretability,  self-interpretable DNNs  have been proposed \citep{ji2025comprehensive}, which integrate interpretability directly into the model architecture or learning objective. Typical approaches include (1) designing sparse or disentangled representations \citep{lemhadri2021lassonet, chen2021nonlinear, luo2023sparse, wang2024penalized, lemhadri2021lassonet,chen2021nonlinear, li2016deep, lu2018deeppink}, (2) imposing structural constraints such as modular layers or attention mechanisms \citep{liu2024kan, vaswani2017attention}, and (3) introducing interpretable intermediate components, for example concept bottlenecks or prototype layers \citep{koh2020concept, li2018deep}. By embedding interpretability into the network, these models allow explanations to emerge naturally from the learned representations rather than relying on external surrogate models.
Despite their advantages, existing self-interpretable DNNs still face several challenges.
First, approaches that impose architectural constraints or introduce interpretable intermediate components often achieve interpretability at the expense of predictive accuracy or model flexibility, as such designs may restrict the network's expressive power \citep{yuksekgonul2022post}.
Second, many methods, such as those based on attention weights, focus on local interpretability without providing a coherent global understanding of the model's decision process \citep{jain2019attention}.
Third, methods based on sparsity assumptions for variable selection assume that only a few strong-signal covariates influence the response.  In practice, however, the effects of individual variables are often weak, and variables tend to influence the response through joint or synergistic effects. Under sparsity assumptions, such joint influential patterns may be missed, potentially leading to degraded model performance.

Finally, both post-hoc interpretability and self-interpretability are often achieved heuristically rather than through statistically principled mechanisms,  none of the existing DNN-based methods  conduct significance testing on individual input variables or their combined representations, which limits their applicability in fields such as medicine and policy-making, where rigorous evidence is essential \citep{lemhadri2021lassonet, farrell2021deep}.

Our key strategy for addressing the aforementioned challenges is to incorporate a learnable $d\times d$ matrix $\bB$ into standard DNNs, where $d$ is the dimensionality of the input $\bX$. Specifically, we adopt the following  \textbf{Deep In}terpretable neural network (DeepIn)
framework:
\begin{equation} \label{obj:proposed}
   Y\sim g(\bB \bX).
\end{equation}
where $g$ is approximated by DNNs.
Unlike classical index models that pre-specify a low-dimensional bottleneck, the DeepIn framework allows $\bB$ to be of size up to $d \times d$, so that the identity matrix is included as a special case. Consequently, when $\bB$ equals the identity matrix, the proposed framework~(\ref{obj:proposed}) reduces to the general model $Y \sim g(\bX)$, imposing no structural restriction on the relationship between $\bX$ and $Y$. Therefore, the full representational capacity of standard DNNs is preserved in (\ref{obj:proposed}).
Importantly, the matrix $\bB$ enables simultaneous feature identification and representation learning by explicitly capturing effect patterns among high-dimensional covariates. In particular, the representation $\bB \bX$ admits an interpretation analogous to that of a linear model, providing explicit expressions of how $\bX$ influences $Y$. Moreover, by identifying and excluding the zero rows of $\bB$, the effective dimension of the representation $\bB \bX$ can be determined automatically, yielding low-dimensional features derived from $\bX$. Variables corresponding to zero columns $\bB_{[,j]}$ are simultaneously excluded, thereby achieving variable selection.

As a result, by imposing the group LASSO penalties
on both the rows and columns of $\bB$, we can implement low-dimensional representation learning and variable selection automatically.
Meanwhile, we incorporate architecture-aware penalty into neural networks for $g$, which can adaptively compress the network based on the learned representation dimension, thereby significantly reducing statistical error while retaining the approximation error, as detailed in Section \ref{sec:method}. 
Finally, we develop a novel hypothesis testing method to examine the significance of the selected variables and the learned  representations.   Rigorous theoretical guarantees for the proposed test are provided in Theorems \ref{thm:normality} and \ref{thm:hypo}. 

Extensive experiments on simulated datasets and four real datasets, including ADNI, BlogFeedback, MNIST, and FashionMNIST,  demonstrate the superior effectiveness of our method in representation learning, variable selection, and network identification, leading to significant improvements in both accuracy and stability, as shown in Tables \ref{table:simulation}, \ref{table:result_reg},  \ref{table:result_mnist} and Supplementary Tables G1 - G3. In particular, compared with off-the-shelf methods such as DFS \citep{chen2021nonlinear}, LassoNet \citep{lemhadri2021lassonet}, and GCRNet \citep{luo2023sparse}, our approach reduces predictive  error by 30.26\% on ADNI and 19.65\% on BlogFeedback. 
Additionally, as shown in Figures \ref{fig:variable_selection_mnist} and \ref{fig:variable_selection_fashion_mnist}, our method automatically identifies human-intuitive discriminative patterns, such as characteristic strokes in digit recognition and garment-specific regions in fashion classification, thereby achieving  model interpretability.
Finally, as shown in Figures \ref{fig:hypothesis} and \ref{fig:hypothesis_combine}, and Supplementary Figures G1 and G3, our hypothesis testing procedure reliably distinguishes informative variables and  representations from noise, achieving near-perfect power for informative variables and over 0.8 for the learned representations, while maintaining the nominal significance level.

In theory, we establish the selection consistency of our estimators. This property provides theoretical guarantees for the identification of important variables, the determination of the minimal number of sufficient representations required to minimize generalization error, and the discovery of efficient sub-network architectures. The selection consistency holds in high-dimensional settings where $\log d = o(n^\alpha)$ for some $0 < \alpha < 1$, simultaneously allowing the number of original network parameters to grow exponentially with the sample size, and the number of selected parameters to diverge as $n \to \infty$. These results ensure effective recovery of both informative inputs and representations in high-dimensional regimes without the need for explicit architectural tuning.
Furthermore, we derive optimal non-asymptotic error bounds for the estimates of the function $g$,  which has the order of $\hat d^{\beta\vee 1+\lfloor\beta\rfloor}(n/\log^c n)^{-2\beta/(\hat d+2\beta)}$, where $\hat d$ is the learned representation dimension.
This finding underscores the critical importance of selecting low-dimensional  representations. 
Building on the established selection consistency and non-asymptotic error bounds, we prove the asymptotic normality of the representation matrix estimator and the functional asymptotic normality of the estimated  function. These results hold even when the number of selected  representations and important variables diverges. Together, they form the theoretical foundation for constructing rigorous statistical inference to test the effects of both  selected variables and learned representations.

The remainder of this paper is organized as follows. Section \ref{sec:method} introduces the proposed method in detail. Theoretical properties of the estimators, including the non-asymptotic error bound and the selection  consistency 
are established in Section \ref{sec:theory}. In Section \ref{sec:hypothesis}, we develop a novel hypothesis testing procedure to assess the significance of variables and representations. The implementation details are provided in Section \ref{sec:imple}. Section \ref{sec:simulation} presents extensive simulation studies, and Section \ref{sec:real_data} applies our method to the ADNI, BlogFeedback, MNIST, and FashionMNIST datasets. Finally, Section \ref{sec:conclusion} concludes the paper. Additional technical proofs and numerical results are included in the Supplementary Materials.

\textbf{Notations.} Let $\beta=q+r$, $r\in (0,1]$ and $q = \left \lfloor \beta \right \rfloor \in \mathbb{N}_0$, where $\left \lfloor \beta \right \rfloor$ denotes the largest integer strictly smaller than $\beta$ and $\mathbb{N}_0$ represents the set of non-negative integers. For a finite constant $B>0$, the
H\"older class $\mcH_\beta([0,1]^d,B)$ 
 is defined as
\begin{eqnarray*}
        \mcH_\beta([0,1]^d,B)=\bigg\{g:[0,1]^d \mapsto \mathbb{R},
        \max_{|\balpha| < q} \| D^{\balpha} g \|_{\infty} \leq B, \max_{|\balpha| = q} \sup_{\bx \neq \by} \frac{\left | D^{\balpha} g(\bx) - D^{\balpha} g(\by) \right |}{\|\bx-\by\|_2^r} \leq B\bigg\},
    \end{eqnarray*}
where  $D^{\balpha} g=\partial^{\balpha} g/\partial x_1^{\alpha_1}\ldots\partial x_d^{\alpha_d}$ denotes the mixed partial derivative of order
$\balpha=(\alpha_1,\ldots,\alpha_d)^{T} \in \mathbb{N}_0^d$ with $|\balpha| = \sum_{i=1}^d\alpha_i$.
The target function $g$ belongs to $\mcH_\beta([0,1]^d,B)$ for given $\beta>0$ and $B>0$.
For $v\in [1,\infty)$, we define the $L^v$-norms $\|g\|_{L^v([0,1]^d)}^v=\int_{[0,1]^d} g^v(\bx) d\bx$, the $v$-norm $\|g\|_v^v=\int g^v(\bx)dP(\bx)$,
and the supermum norm $\|g(\bx)\|_{L^\infty([0,1]^d)}=\|g\|_\infty=\sup\limits_{\bx} |g(\bx)|$. The same notation $\|\cdot\|_v$ applies to the $v$-norm of vectors and matrices.

\section{Method}
\label{sec:method}

Theoretical analysis shows that our statistical inference, detailed in Section \ref{sec:hypothesis},
involves estimating both the target functions and their derivatives. Derivative estimation requires the use of deep neural networks (DNNs) with Rectified Power Unit (RePU) activations of order $p\geq 2$, as established in Supplementary Corollary 2.
Accordingly,  we consider  a class of RePU-activated
neural networks  $\mcG$ to approximate the target function, 
where each $g_n \in \mcG$ takes the form $
g_n(\bx) = \bW^\top_\mcD\sigma\left(\bW^\top_{\mcD-1}\sigma(\bW^\top_{\mcD-2}\sigma(\bW^\top_{\mcD-3} \ldots \sigma(\bW^\top_{0}\bx + \ba_0)) + \ba_{\mcD-2}) + \ba_{\mcD-1}\right) + a_\mcD,
$
with $\sigma(x)= \{\max(x, 0)\}^p$ denoting  the RePU activation,  applied elementwise for $p\geq 2$. 
The network architecture is characterized by  \textit{depth} $\mathcal{D}$, \textit{width} $\mathcal{W}$, \textit{size} $\mathcal{S}$, and \textit{number of neurons} $\mathcal{U}$.
Let $d_j$ denote the width of layer $j$ for  $j=1, \cdots,  \mcD+1$, where the input and output dimensions are  $d$ and $d_{\mathcal{D}+1} = 1$. 
Then $\mathcal{W}=\max\{d_1, d_2, \ldots, d_\mathcal{D}\}$,
$\mathcal{S}=\sum_{i=0}^{\mathcal{D}} d_{i+1} (d_i+1)$, and $\mathcal{U} = \sum_{i=1}^{\mathcal{D}} d_{i}$.

Denote by $\btheta=\{(\bW_{j}, \ba_{j}), j = 0, \ldots, \mcD\}$  the collection of network weights. Accordingly, we parametrize  $g_n\in \mcG$ as $g_{\btheta}$, and define  the estimation space as $\mcG_{\bB}:=\mcG\times\mathbb R^{d\times d}$ for  $(g,\bB)$ defined in model (\ref{obj:proposed}). 
To automatically perform representation learning  and variable selection, we adopt a group lasso penalty on both the rows and columns of $\bB$. Furthermore, based on the number of learned representations, we incorporate an additional network penalty on $g_{\btheta}$ to ensure that the network architecture aligns with the identified representation dimensionality.
Then the estimator for the parameter $\bvmu:=(\btheta,\bB)$ is obtained as
\begin{eqnarray}
\label{eq:group}
\hat \bvmu =\mathop{\arg\min}\limits_{\bvmu: (g_{\btheta},\bB)\in \mcG_{\bB}} \mcR_n(g_{\btheta},\bB):=\mathop{\arg\min}\limits_{\bvmu: (g_{\btheta},\bB)\in \mcG_{\bB}}
 \frac{1}{n}\sum\limits_{i=1}^n L(g_{\btheta}, \bB; \bS_i) 
+ \blambda \brho(\bvmu),
\end{eqnarray}
where  $L(g_{\btheta}, \bB; \bS_i)\hat=\tilde L(g_{\btheta}(\bB\bX_i), Y_i)$, $\tilde L(\cdot,\cdot)$ is a loss function, such as the least squared (LS) loss for regression or  the cross-entropy loss for classification,
$\bS_i=(\bX_i, Y_i)$,  
$\brho(\bvmu)=(\rho_1(\bvmu),\rho_2(\bvmu),\rho_3(\bvmu))
=\Big(\sum_{k=1}^{d}\|\bB_{[k,]}\|_{2},
\sum_{k=1}^d\|\bB_{[,k]}\|_2, \|\btheta\|_1\Big)$, $\blambda \brho(\bvmu)=\sum\limits_{j=1}^3\lambda_j\rho_j(\bvmu)$,
and $\bB_{[k,]}$ and $\bB_{[,k]}$ denote the $k$-th row and $k$-column of the matrix $\bB$ respectively.  Here the first penalty identifies the number of  representations by detecting and excluding zero rows of $\bB$. Since column $j$ of $\bB$ represents the effect of variable $j$, a zero column indicates that variable $j$ has no impact on $Y$. Therefore, the second penalty on columns is used for  variable selection, while the third penalty controls the network complexity to match the effective dimension of the function $g$, i.e., the number of selected representations.
Further details regarding $\rho_3(\bvmu)$ and the implementation of the proposed method are provided in Section~\ref{sec:imple}. 

\section{Selection consistency and  error analysis}
\label{sec:theory}



Throughout the paper, the subscript ``$0$" is used to denote the true value or the corresponding minimal quantity.
Due to the inherent unidentifiability of neural networks, we define the  parameter set required to achieve the smallest approximation error by
$\Upsilon_0:=\{\bvmu: \bvmu\in \mathop{\arg\min}\limits_{\bvmu: (g_{\btheta},\bB)\in \mcG_{\bB}}\|g_{\btheta}(\bB\bX)-g_0(\bB_0\bX)\|_{L^v([0,1]^d)}\}$  for some
$v\in [1,\infty]$.  Because multiple representation dimensions may be capable of approximating the target function
$g_0(\bB_0\bX)$, we introduce the following definition of the minimal representation dimension.

\begin{dfn} \label{def:mfd}
The minimal representation  dimension (MRD) is defined as
$d_0 := \underset{d}{\min}\{d \ | \ \bvmu =(\btheta,\bB) \in \Upsilon_0, \ \mbox{where the number of nonzero rows of} \ \bB \  \mbox{is} \ d \}.$
\end{dfn}

Without loss of generality, we assume that only the first $d_0$ rows and the first $s_0$ columns of $\bB_0 \in \mathbb{R}^{d\times d}$ are nonzero, while all other entries are zero,
where 
$s_0$ is  the number of important variable in $\bX$.
We further introduce $\mcS_0^{net}:=\min\limits_{\bvmu\in\Upsilon_0}\|\btheta\|_{0}$ as the minimal network size required to achieve optimal approximation.
Denote $\mcS_0:=(d_0,s_0,\mcS^{net}_{0})$, and  
let $\Upsilon^{min}=\{\bvmu:\bvmu\in\Upsilon_0, \mcS_{\mu}=\mcS_0\}$ be the set of parameters achieving the smallest approximation error with a minimal network under the MRD framework.
Then the minimal number of parameters required to achieve the smallest approximation error is $\mcT_0 =\mcS_0^{net}+d_0s_0$. We regard  $\bvmu_0=(\btheta_0,\bB_0)\in\Upsilon^{min}$ as the true values.

\begin{dfn}[Selection consistency]
\label{def:net}
An estimator $\hat{\bvmu}$ achieves selection consistency if
$P(\mcS_{\hat\bvmu}= \mcS_0) \to
 1$ as $n \to \infty$, where
 $\mcS_{\bvmu}:=\Big(\sum\limits_{k=1}^d I(\|\bB_{[k,]}\|\neq 0), \sum\limits_{k=1}^d I(\|\bB_{[,k]}\|\neq 0), \|\btheta\|_{0}\Big)$ denotes the triplet consisting of  the number of active  representations, the number of important variables, and the number of nonzero neural network weights.
\end{dfn}

To establish the selection consistency, we define the oracle estimation class
$$
\Upsilon^{org}:=\{\bvmu: \bvmu\in \underset{\bvmu: (g_{\btheta},\bB) \in \mcG_{\bB}}{\arg \min}\mcR_n(g_{\btheta},\bB), \mcS_{\bvmu} = \mcS_0\},
$$
where $\mcS_{\bvmu}=\mcS_0$  implies that  $(d_0,s_0,\mcS^{net}_{0})$ are known.
If we can show that $P(\hat\bvmu\in \Upsilon^{org})\to 1$, then selection consistency is established. 

According the definition of $\bB_{0}$,   the active index sets for non-zero  representations and  important variables are  $\mcA_1=\{1,\cdots, d_0\}$  
and  $\mcA_2=\{1,\cdots, s_0\}$, respectively. 
The index set of non-zero network weights for $\btheta_0$ within a minimal network is defined as $\mcA_3(\btheta_0)=\{k: \theta_{0k}\neq 0, \bvmu_0=(\btheta_0,\bB_0)\in\Upsilon^{min}\}$, where  $\theta_{0k}$ denotes  the $k$-th element of the  vectorized   $\btheta_0$. 
The half minimal signal strength among these effects is then
 $l_n=\min\big\{\min\limits_{k\in \mcA_1} \|\bB_{0[k,]}\|_2,  \min\limits_{k\in \mcA_2} \|\bB_{0[,k]}\|_2,  \min\limits_{k\in \mcA_3(\btheta_0)}|\theta_{0k}|\big\}/2$.
 As shown in Supplementary Proposition C2, we denote the approximation error as $\eta_{net}:= c_1Bd_0^{(\beta\vee 1+\lfloor\beta\rfloor)/2}(MN)^{-2\beta/d_0}$, where $M$ and $N$ are positive integers related to depth and width respectively, and $c_1$ is a constant depending on $\beta$.

\begin{thm}[Selection consistency]\label{thm:selection}
For some $0\leq\nu_2<\nu_1\leq 1/2$, define $\Delta_{net,n}:=Cn^{-1/2}+D\log n\cdot \eta_{net}^2 \mcT_0^{1-2\nu_2}$, where $C$ and $D$ are positive constants.
Under Supplementary Assumptions 1-5,
if for some constants $C_1, C_2$ and $C$,
\begin{align}\label{eq:6}
\nonumber
&
 \min\big\{\blambda\circ\brho_m(\hat\bvmu_{\mcA^c})\big\}
 \geq I\big(\min\big\{\brho_m(\hat\bvmu_{\mcA^c})\big\}\neq 0\big)\Delta_{net,n},\ \ \
 C_1\eta_{net}/\mcT_0^{\nu_1}\leq\lambda_m\leq C_2\eta_{net}/\mcT_0^{\nu_2},
 \\
&
 l_n>C\lambda_m\mcT_0, \ \  \  \
N(\delta_1, \delta_2)\sum\limits_{j=1}^3\exp(-n r_{jn}^2 t^2)+\exp(-Cn\eta_{net}^2)\to 0,
\end{align}
where
$\lambda_m:=\lambda_1\vee\lambda_2\vee\lambda_3$ with $\lambda_j=R_j(t)$,
$\brho_m(\hat\bvmu_{\mcA^c}):=\Big(\sum\limits_{k\in\mcA_1^c}\|\hat\bB_{[k,]}\|_2, \sum\limits_{k\in\mcA_2^c}\|\hat\bB_{[,k]}\|_2, $ $
\min\limits_{k\in\mcA_3^c(\btheta_0)}|\hat\theta_k|
\Big)$, 
the $\circ$ denotes the elementwise multiplication operator,
and $N(\delta_1,\delta_2)$, $r_{jn}$ and $R_j(t)$ are defined in Supplementary,
then
$P(\hat \bvmu\in\Upsilon^{org})\to 1$ as $n\to \infty$.
\end{thm}

In (\ref{eq:6}),
$\Delta_{net,n}$ represents an upper bound on the empirical excess risk between the estimated network $\hat\bvmu\in \mathbb R^{d_{\bvmu}}$ and its projection $\tilde\bvmu$ onto a minimal network.
Specifically, we have
$|\mathbb P_n(L(g_{\hat\btheta},\hat\bB;\bS)-L(g_{\tilde \btheta},\tilde\bB;\bS))|\leq \Delta_{net,n}$.
When the elements of tuning parameter $\blambda$ are large enough so that $\min\big\{\blambda\circ\brho_m(\hat\bvmu_{\mcA^c})\big\}
 \geq I\big(\min\big\{\brho_m(\hat\bvmu_{\mcA^c})\big\}\neq 0\big)\Delta_{net,n}$,
the redundant network parameters, irrelevant representations, and unimportant variables of $\bX$ can be  effectively eliminated.

The second condition $C_1\eta_{net}/\mcT_0^{\nu_1}\leq\lambda_m\leq C_2\eta_{net}/\mcT_0^{\nu_2}$ in (\ref{eq:6}) requires the
tuning parameter $\lambda_m$ to be of order $O(\eta_{net}/\mcT_0^{\nu})$ for some $\nu\in(\nu_2,\nu_1)$.
The third condition  implies that
the minimum signal strength must exceed the threshold $\lambda_m\mcT_0$, which depends on both the minimal size $\mcT_0$ and the choiced tuning parameter $\lambda_m$. These conditions ensure the identifiability of significant representations, important variables and  sub-networks of minimal size.

When $N(\delta_1,\delta_2)$ is finite, the final condition in (\ref{eq:6}) is satisfied provided that  $1/(r_{jn}^2t^2)=o(n)$ for $j=1,2,3$ and $1/\eta_{net}=o(\sqrt{n})$.  By the definition of $r_{jn}$ in Supplementary,  the condition $1/(r_{jn}^2t^2)=o(n)$ can be satisfied when the covariate dimension $d$ grows at an exponential rate and the dimension of selected variables $s_0$ grows polynomially with sample size, given an appropriately chosen tuning parameter $t$. 
The second condition $1/\eta_{net}=o(\sqrt{n})$ requires  the approximation error $\eta_{net}$ to decay more slowly than $n^{-1/2}$. Thus, Theorem \ref{thm:selection} demonstrates that with suitable $\blambda$ and sufficiently strong signals in both the network and representation  matrix, we can achieve effective signal recovery, ensuring  selection consistency with high probability.

Next, we establish both estimation consistency and non-asymptotic error bounds.  

\begin{thm}[Non-asymptotic error bound] \label{thm:prediction_bound}
For integers $M$ and $N$, we can obtain a RePU DNN estimator
$(g_{\hat\btheta},\hat\bB)$ with width at most
$p(\lfloor\beta\rfloor+1)^2 d^{\lfloor\beta\rfloor+1}N\lceil \log_2(4N)\rceil$,
depth $\mcD$ at most
$13M\lceil \log_2(4M)\rceil+\max\{0, \lceil\log_p(\lfloor\beta\rfloor)\rceil-1\}$,
and size at most
$184pN^2M(\lfloor\beta\rfloor+1)^4d^{\lfloor\beta\rfloor+1}\lceil \log_2(8NM)\rceil$.
Under the conditions of Theorem \ref{thm:selection},
suppose that  
\begin{equation}\label{eq:7}
\small{Bd_0^{\beta\vee 1+\lfloor\beta\rfloor}(NM)^{-4\beta/d_0}\to 0,    \  \
\iota_{net,n}\to 0}, \ \  \mbox{and} \ \  \min_{\bvmu_0\in\Upsilon^{min}}\blambda \brho(\bvmu_0)\leq \iota_{net,n}/2,
\end{equation}
where the $\iota_{net,n}$ is stochastic error,  defined as
$$\iota_{net,n}:=3^2 2^{24}\max\{B_{7n}^2,4B_8^2\}\mcD^2(\log n)\mcT_0\log(p\mcT_0)/n$$
with $B_{7n}$ and $B_8$ being defined as in Supplementary Assumption 5.
Then\\
\noindent(1) (Estimation consistency) The estimator satisfies
$E\|g_{\hat\btheta}(\hat\bB\bX)-g_0(\bB_0\bX)\|_2^2\to 0$.

\noindent(2) (Non-asymptotic error bound)
The error bound  for large $n$ satisfies
\begin{align*}
E\|g_{\hat\btheta}(\hat\bB\bX)-g_0(\bB_0\bX)\|_2^2
\leq
\frac{2c_1^2 B_5}{B_4}B^2 d_0^{\beta\vee 1+\lfloor\beta\rfloor}(NM)^{-4\beta/d_0}+\frac{3\iota_{n,net}}{B_4}.
\end{align*}
\end{thm}

The first condition in (\ref{eq:7}) requires the original network to be sufficiently large, while the second condition necessitates proper control of the minimal size $\mcT_0$ and the original depth $\mcD$.
When the penalty term is properly controlled, the third condition
$\min_{\bvmu_0\in\Upsilon^{min}}2\blambda \rho(\bvmu_0) \leq \iota_{net,n}$ holds.

The  error bound in Theorem \ref{thm:prediction_bound} decomposes into two components: an approximation error of order $d_0^{\beta\vee 1+\lfloor\beta\rfloor}(NM)^{-4\beta/d_0}$,  and the stochastic error $\iota_{net,n}$ of order $\max\{B_{7n}^2,4B_8^2\}(\log n)\mcD^2\mcT_0 \log (p\mcT_0)/n$.
Theorem \ref{thm:prediction_bound} reveals the approximation error is driven by the MRD  $d_0$, while  the stochastic error scales with $\mcD^2\mcT_0$. These results 
emphasize the importance of selecting the MRD and a minimal  network that achieves the optimal approximation error, as these factors directly determine the generalization performance.

\begin{corollary}[Optimal non-asymptotic error bound]\label{coro:WFD}
Under the conditions of Theorem \ref{thm:prediction_bound}, when $N=(n/\log^c n)^{d_0/2(d_0+2\beta)}$ and $M$ is a finite constant, the DNN estimator achieves the optimal error bound
$$
E\|g_{\hat\btheta}(\hat\bB\bX)-g_0(\bB_0\bX)\|_2^2
\leq C B^2 d_0^{\beta\vee 1+\lfloor\beta\rfloor}(n/\log^c n)^{-2\beta/(d_0+2\beta)}
$$
for sufficiently large $n$, where $C$ is a positive constant.  The exponent $c$ equals $3$ when $B_{7n}$ is finite and $5$ when $B_{7n}$ is divergent.
\end{corollary}

Corollary \ref{coro:WFD} implies the optimal non-asymptotic error bound is achieved by a  network with a fixed-depth,  width  not exceeding
$38p(\lfloor\beta\rfloor+1)^2d_0^{\lfloor\beta\rfloor+1}
\lceil n_c^{d_0/2(d_0+2\beta)}\log_2\{4n_c^{d_0/2(d_0+2\beta)}\}\rceil$
and  size  bounded by
$2(\lfloor\beta\rfloor+1)^4d_0^{\lfloor\beta\rfloor+1}\lceil
\log_2\{8Mn_c^{d_0/2(d_0+2\beta)}\}\rceil\{52Mn_c^{d_0/(d_0+2\beta)}$ $+40pM\}$, where $n_c=n/\log^c n$.
The term $B_{7n}$ is determined by the model and the loss function. For instance, $B_{7n}=O(\log n)$ for least squares loss in a regression model with an unbounded response, while it is finite for the negative log-likelihood loss in logistic regression.


While data are often observed in a high-dimensional ambient space, the essential variation of the target function typically lies on a much lower-dimensional structure. Manifold learning theory \citep{Belkin2004,Fefferman2016} formalizes this idea, showing that the  complexity of learning is determined by the intrinsic, rather than ambient, dimension. Particularly, when the predictor $\bX$  is supported on $\mcM_\rho$, a neighborhood of a compact $d_{\mcM}$-dimensional Riemannian manifold $\mcM\subset [0,1]^d$ with condition number $1/\tau$ \citep{Hubbard2015},  the convergence rates adapt to the intrinsic manifold dimension. For instance, Theorem 6.2 in \cite{jiao2023deep} shows that under exact manifold assumptions, the ReLU DNN estimator satisfies
\begin{eqnarray}
\label{manifold}
E|g_{\hat\btheta}(\bX)-g_0(\bX)|^2
\leq
CB^5(\lfloor\beta\rfloor+1)^9(6/\tau)^{2d_{\mcM}}d_{\mcM}^{3\lfloor\beta\rfloor+6}d(\log n)^8 n^{-2\beta/(d_{\mcM}+2\beta)}.
\end{eqnarray}
The manifold rate $n^{-2\beta/(d_{\mcM}+2\beta)}$ and the optimal rate $n^{-2\beta/(d_0+2\beta)}$ established in Corollary \ref{coro:WFD} exhibit similar convergence behavior, indicating that the selected dimension $d_0$ effectively serves as the intrinsic dimension $d_{\mcM}$. An important distinction, however, is that the error  bound under manifold assumptions still scales linearly with the ambient dimension $d$, our optimal rate depends solely on the identified intrinsic dimension $d_0$, which is typically much smaller than $d$. This highlights the effective dimension reduction achieved by our estimators and demonstrates that our method can efficiently handle high-dimensional problem.


In addition, the manifold rate  alone provides limited guidance in practice. For example, as seen in Tables \ref{table:simulation} and Supplementary  G1-G3, even when the underlying network has a low-dimensional structure, conventional DNN training yields  considerably lower accuracy than our method that fully explores and exploits the intrinsic  dimensionality.
The main reason is that, without prior knowledge of task complexity, network architectures are typically selected heuristically or with only limited manual tuning, making it difficult to achieve the theoretical optimum.
In contrast, our method adapts to the intrinsic geometry of the data and thus achieves optimal performance in both theory and practice. It therefore bridges the gap between minimax theory and practical implementation.

\section{Hypothesis testing}
\label{sec:hypothesis}

Our analysis begins in the oracle space $\mcG_{\bB}^{org}=\{(g_{\btheta},\bB): (g_{\btheta},\bB)\in\mcG_{\bB}, \mcS_{\bvmu}=\mcS_0\}$, where we establish the asymptotic normality of $\hat\bB$ and the functional normality of $g_{\hat\btheta}$. Theorem \ref{thm:selection}, which guarantees selection consistency, allows us to generalize these results to the full  space $\mcG_{\bB}$. Therefore, for notational simplicity, we hereafter let $\bX\equiv\bX_{\mcA_2}$ represent the active set of predictors and, with  slight abuse of notation,  take $\mcG_{\bB}$ to be the oracle subspace $\mcG\times \mbR^{d_0\times s_0}$.
Since the MRD $d_0$ is typically smaller than the number of covariates, we assume that the representation extractor $\bB \in \mbR^{d_0\times s_0}$ has full row rank.

\subsection{Testing for individual covariate effects}

$\bB_{0[,j]}=0$ implies that covariate $j$ has no effect on $Y$. Hence, significant variables in $\bX$ can be identified by testing the hypotheses
$H_0: \|\bB_{0[,j]}\|_2=0 \ \mbox{v.s.} \  H_1: \|\bB_{0[,j]}\|_2\neq 0$ for each predictor component $j=1,\ldots,s_0$.
However, since $\bB$ in the model \eqref{obj:proposed} is not identifiable, directly testing these hypotheses is  infeasible. To overcome this issue, we introduce an identifiable representation matrix $\tilde\bB_0 \in \mathbb{R}^{d_0\times s_0}$ satisfying the following condition:
\begin{con}\label{assumption5}
The matrix $\tilde\bB_0$ has full row rank and satisfies $g_0(\bB_0\bX)=\tilde g_0(\tilde\bB_0\bX)$ for any $\bX$. Furthermore, $\tilde\bB_0$ is constrained such that (i) the first nonzero element in each row is strictly positive, (ii) each row is normalized to have unit $\ell_2$-norm, i.e., $\|\tilde\bB_{0[j,]}\|_2=1$, and (iii) its singular value decomposition (SVD) produces a sequence of positive and monotonically decreasing singular values.
\end{con}

Following \citet{Yuan2011}, it can be shown that $\tilde\bB_0$ is identifiable. We estimate $\tilde\bB_0$ using the proposed method with an additional normalization step to ensure that the resulting estimator $\hat{\tilde\bB}$ satisfies Condition \ref{assumption5}.
Define $L^{(l)}(u;\bS)=\partial^l L(u;\bS)/\partial u^l$ for $l\ge 1$.
Let $\Lambda_1=\Lambda_{\max}(\bJ^\top(\tilde\bB_0)\bV_1^{-2}\bJ(\tilde\bB_0))$, $\Lambda_2=\Lambda_{\max}(\bV_3)$ and $\Lambda_3=\Lambda_{\min}(\bV_3)$, where $\Lambda_{\max}$ and $\Lambda_{\min}$ denote the largest and smallest eigenvalues of a matrix respectively. The matrices $\bJ(\tilde\bB_0)$ and $\bV_k, k=1,2,3$ are defined in Supplementary A.
In Theorem \ref{thm:normality}, we first establish the asymptotic normality of $\hat{\tilde\bB}$, and then demonstrate the asymptotic equivalence of test statistics constructed from $\hat{\bB}$ and $\hat{\tilde\bB}$.
These results provide a theoretical foundation for testing the  significance of each predictor $\bX_{[j]}$ using any  estimator for $\bB$ obtained from (\ref{eq:group}).

\begin{thm}\label{thm:normality}
Let $\alpha_n=d_0^\beta (n/\log^c n)^{-(\beta-1)/(d_0+2(\beta-1))}$, 
where  $c$ is the constant from Corollary \ref{coro:WFD}.
Under the assumptions of  Corollary \ref{coro:WFD} and condition (D3) in Supplementary D with $r\geq 1$, suppose the Lipschitz condition holds for the $l$-th derivative of $L(\cdot,\cdot)$ for  $l\geq 1$, namely,
 \begin{equation}\label{eq:lipshitz}
 |L^{(l)}(g_{\btheta_1},\bB_1;\bS)-L^{(l)}(g_{\btheta_2}, \bB_2;\bS)|\leq C_S |g_{\btheta_1}(B_1\bX)-g_{\btheta_2}(B_2\bX)|,
 \end{equation}
where $C_S$ may depend on $\bS$ with $\mbE e^{aC_S}<\infty$ for some constant $a>0$.
Moreover, we suppose that $\mbE [L^{(1)}(\tilde g_0,\tilde\bB_0;\bS)|\bB_0\bX]=0$
and  $\mbE [L^{(1)2}(\tilde g_0,\tilde \bB_0;\bS)|\bB_0\bX]=\varsigma_1^2$.
Further, we assume
\begin{align}\label{eq:normcond}
\frac{d_0s_0\Lambda_1}{n}\to 0, \
\frac{nd_0s_0\Lambda_1}{\Lambda_2}\alpha_n^4\to 0, \
\frac{\Lambda_1\Lambda_2}{n}\to 0, \
\frac{d_0s_0}{n\Lambda_3^2}\to 0, \
\frac{n(d_0+s_0)(\lambda_1\vee\lambda_2)^2}{\Lambda_2\wedge (\Lambda_3/\Lambda_1)}\to 0.
\end{align}
Then for $\beta>1$:

\noindent
(1) (Asymptotic normality)
$\sqrt{n}\bu^\top\bSigma^{-1}\big(\hat{\vec{\tilde\bB}}-\vec{\tilde\bB}_0\big)\xrightarrow{d} N(0, 1)$. Here $\bu$ is any unit vector in
$\mbR^{d_0(s_0-1)}$,
$\vec{\tilde\bB}$ is the vectorization of $\tilde\bB^\top$,
and $\bSigma^{-1}=\varsigma_1^{-1}\bV_3^{-1/2}$.

\noindent
(2) (Hypothesis testing)
Under the null hypothesis $H_0: \|\bB_{0[,j]}\|_2 = 0$,
for any unit vector $\bu_j\in\mbR^{d_0}$,
$$\sqrt{n}\varsigma^{-1}\bu_j^\top\big(\bar\bV_3^{-1/2}\hat{\vec\bB}\big)_{[(1:d_0)s_0+j-s_0]}=\sqrt{n}\varsigma^{-1}\bu_j^\top\big(\bV_3^{-1/2}\hat{\vec{\tilde\bB}}\big)_{[(1:d_0)s_0+j-s_0]},$$
where $[(1:d_0)s_0+j-s_0]$ indexes the vector corresponding to $\bB_{0[,j]}$, and $\bar \bV_3$ is defined in Supplementary A.
\end{thm}


When $\Lambda_j, j=1,2,3$ are finite, condition (\ref{eq:normcond}) holds if $d_0s_0 = o(n)$, $d_0s_0\alpha_n^4=o(n^{-1})$ and $\lambda_1 \vee \lambda_2$ is sufficiently small. These requirements are  satisfied, for instance, if $d_0s_0 = O(n^{\alpha_4})$ and $\alpha_n=o(n^{-(1+\alpha_4)/4})$ for some $\alpha_4 < 1$. Consequently, asymptotic normality and valid hypothesis testing remain attainable even when the intrinsic dimension  $d_0$  and the sparsity level $s_0$ diverge.

Furthermore, under the null hypothesis, the equivalence of statistics related to $\hat\bB$ and $\hat{\tilde\bB}$ implies that $\sqrt{n}\varsigma^{-1}\bu_j^\top\big(\bar\bV_3^{-1/2}\hat{\vec\bB}\big)_{[(1:d_0)s_0+j-s_0]}\xrightarrow{d} N(0, 1)$.
Consequently, the following testing procedure remains valid within the estimation space $\mcG_{\bB}$  even when the true parameter $\bB_0$ is not uniquely identifiable.
\begin{itemize}
\spacingset{1}
\item The conditional expectation $E(\bX|\bB_0\bX)$ is estimated nonparametrically via a neural network, where the unknown matrix $\bB_0$ is replaced by $\hat\bB$.

\item The matrices $\hat\bV_1$ and $\hat\bV_2$ are obtained by replacing $\bB_0$ and $g^{(1)}_0$ with their respective estimators, $\hat\bB$ and $g^{(1)}_{\hat\btheta}$, in the empirical counterparts of $\bV_1$ and $\bV_2$.

\item We estimate the error variance as $\hat\varsigma_1^2=\mathbb P_n L^{(1)2}(g_{\hat\btheta},\hat\bB;\bS)$, and obtain $\hat{\bar\bV}_3=\bar\bJ^\top(\hat\bB)\hat\bV_1^{-1}\hat\bV_2\hat\bV_1^{-1}\bar\bJ(\hat\bB)$.

\item For each predictor component $\bX_{[j]}$, we compute the test statistic
$$\bU^{[j]}=\sqrt{n}\hat\varsigma^{-1}\big(\hat{\bar\bV}_3^{-1/2}\hat{\vec\bB}\big)_{[(1:d_0)s_0+j-s_0]}.$$

\item The corresponding $p$-value is obtained by $P(\chi^2\geq \|\bU^{[j]}\|_2^2)$, where $\chi^2$ follows a $\chi^2$-distribution with $d_0$ degrees of freedom.

\end{itemize}

\subsection{Testing for the effects of combined  representations}

Let $\bz=(z_1,\cdots, z_{d_0})^T := \bB_0\bx$, where $z_i = \bB_{0[i,]}^\top\bx$ denotes the $i$th combination representation in $g_0$ for $i=1,\ldots,d_0$.
For a given index set $\mcI \subseteq [1:d_0]$ with complement $\mcI^c = [1:d_0]\setminus \mcI$, we define $\bz_{\mcI}=(z_{\mcI,1},\cdots, z_{\mcI,d_0})^T$ such that $z_{\mcI,j}=z_j$ for $j\in \mcI$ and $z_{\mcI,j} = 0$ for $j \in \mcI^c$.

If $g_0(\bz) = g_{0\mcI}(\bz_{\mcI})$, then the combination representations $\bB_{0[i,]}^\top \bX$ for $i \in \mcI^c$ are not statistically significant. Hence, testing whether the combination representations $\bB_{0[i,]}^\top \bX$, $i \in \mcI$, have significant effects on $Y$ is equivalent to testing the hypothesis
$$H_0: g_0(\bz)=g_{0\mcI}(\bz_{\mcI}) \ v.s. \  H_1: g_0(\bz)\neq g_{0\mcI}(\bz_{\mcI}),$$
where the test statistics are constructed  based on the following functional asymptotic normality.


\begin{thm}[Functional normality]\label{thm:normality_g}
Under the conditions of Theorem \ref{thm:normality}, assuming (\ref{eq:lipshitz}) holds for $l\geq 2$, then we have
\begin{align*}
\sqrt{n}\int L^{(2)}(g_0,\bB_0;\bs)
\big\{\big[g_{\hat\btheta}(\bB_0\bx)-g_0(\bB_0\bx)\big]
+\big[g_0(\hat\bB\bx)-g_0(\bB_0\bx)\big]\big\}dP(\bs)
\xrightarrow{d} N(0,\varsigma_2^2),
\end{align*}
where
 $P(\bs)=P(\bx,y)$ is the probability measure of $\bS=(\bX,Y)$ and $\varsigma_2^2=\mbE m_1^2(g_0,\bB_0;\bS)$, where $m_1$ is defined in Supplementary A.
\end{thm}

To leverage this theorem for constructing test statistics, we split the dataset $\bS$ into two  independent subsets $\bS_1$ and $\bS_2$, with sample size $n_1$ and $n_2$ respectively. Let $\mathbb P_{n_j}$ denote the empirical measure of $\bS_j$ for $j=1,2$.
For subset $j$, we obtain the estimators
$\hat g_{\bvmu}^{[j]}:=\big(g_{\hat\btheta}^{[j]},\hat\bB^{[j]}\big)$ for $g_0(\bB_0\bx)$, and
$\hat g_{\bvmu_\mcI}^{[j]}:=\big(g_{\hat\btheta_{\mcI}}^{[j]},\hat\bB_{\mcI}^{[j]}\big)$ for $g_{0\mcI}(\bB_{0\mcI}\bx)$.
Here $\bB_{0\mcI}$ is the matrix obtained from $\bB_0$ by keeping its $i$th row if $i \in \mcI$ and replacing it with the zero vector otherwise. We assume that, both $\hat\bB^{[j]}$ and the submatrix of $\hat\bB_{\mcI}^{[j]}$ formed by the rows $i \in \mcI$,  satisfy Condition 1, ensuring that the estimators of $(g_0,\bB_0)$ and $(g_{0\mcI},\bB_{0\mcI})$ are well-defined on both subsets $\bS_1$ and $\bS_2$.
The test statistic for  $H_0: g_0(\bz)=g_{0\mcI}(\bz_{\mcI})$ is constructed as
$$T_n=\sqrt{n}(T_n^{[1]}+T_n^{[2]}),$$
where for $j,k=1,2$ and $k\neq j$,  $T_n^{[j]}:=\frac{n_j}{n}\mathbb P_{n_j}
u(\hat g_{\bvmu_{\mcI}}^{[j]},\hat g_{\bvmu}^{[k]};\bS_j),$ and
$$
u(\hat g_{\bvmu_{\mcI}}^{[j]},\hat g_{\bvmu}^{[k]};\bS_j)
:=
L^{(2)}\Big(\hat g_{\bvmu_{\mcI}}^{[j]};\bS_j\Big)
\Big\{\Big[g_{\hat\btheta}^{[k]}\big(\hat\bB_{\mcI}^{[j]}\bX_j\big)
-g_{\hat\btheta_{\mcI}}^{[j]}\big(\hat\bB_{\mcI}^{[j]}\bX_j\big)\Big]
+\Big[g_{\hat\btheta_{\mcI}}^{[j]}\big(\hat\bB^{[k]}\bX_j\big)
-g_{\hat\btheta_{\mcI}}^{[j]}\big(\hat\bB_{\mcI}^{[j]}\bX_j\big)\Big]\Big\}.
$$
Obviously, $T_n$ is the empirical counterpart of the statistic discussed in Theorem \ref{thm:normality_g}. As a result, we have the following theorem.

\begin{thm}[Asymptotic normality of test statistic]\label{thm:hypo}
Assume the conditions of Theorem \ref{thm:normality_g} hold, and let $n_1/n\to p$ as $n\to \infty$ with $0<p<1$. Then under $H_0: g_0(\bz)=g_{0\mcI}(\bz_{\mcI})$, the following results hold:

(1) The test statistic $T_n$ converges in distribution to $N(0,\varsigma_2^2)$.

(2) The asymptotic variance $\varsigma_2^2$ can be consistently estimated by
$\hat\varsigma_2^2=\sum\limits_{j=1}^2\mathbb P_{n_j}(L^{(1)}(\hat g_{\bvmu_{\mcI}}^{[j]};\bS_j))^2/2$.
\end{thm}

Theorem \ref{thm:hypo} requires the two split datasets to have sample sizes of the same asymptotic order. Based on this result, we construct the test statistic $T_n/\hat\varsigma_2$, which asymptotically follows a standard normal distribution. This provides a rigorous foundation for testing the significance of the combination effects $\bB_{[i,]}^\top\bX$ for $i \in \mcI$, under the null hypothesis $H_0: g_0(\bz) = g_{0\mcI}(\bz_{\mcI})$.

\section{Implementations}
\label{sec:imple}

In this section, we detail the implementation of the proposed method. We use a mini-batch stochastic gradient descent algorithm \citep{bottou2010large} to solve \eqref{eq:group}. Directly optimizing \eqref{eq:group} with gradient-based methods is challenging because the penalty terms $\lVert \cdot \rVert_{2}$ and $\lVert \cdot \rVert_{1}$ are nondifferentiable at zero. To address this issue, we adopt a subgradient descent algorithm \citep{boyd2004convex} combined with parameter truncation. Three thresholding hyperparameters, $\tau_1$, $\tau_2$, and $\tau_3$, are used to truncate rows and columns of $\hat{\bB}$ with small $L_2$ norms, and elements of $\hat{\btheta}$ with small absolute values, respectively. Algorithm \ref{alg:pta} illustrates the truncation procedure for the row-wise operation with threshold $\tau_1$. The same idea applies to column-wise truncation using $\tau_2$ and element-wise truncation of $\btheta$ using $\tau_3$. The complete training pipeline of the proposed method is summarized in Algorithm \ref{alg:all}. Additional details on the subgradient computation and the selection of $\tau_1$, $\tau_2$, and $\tau_3$ are provided in Supplementary E.


There are several technical considerations in implementing the algorithm. First, Corollary \ref{coro:WFD} shows that the optimal error bound is achieved when the network depth is fixed. To maintain good performance, we therefore regulate the network depth when controlling network complexity so that it aligns with the number of selected representations. Specifically, if a layer $l$ is redundant, the affine transformation
$T_l(\bx) = \bW_l \bx + \ba_l$ becomes unnecessary; that is, the information carried by $T_l(\bx)$ is equivalent to that of $\bx$. This implies  $\bW_l$ can be replaced by the identity matrix and $\ba_l=0$, which  motivates the depth penalty  $\rho_{31}(\bvmu)=\sum_{l=1}^{\mcD}\|\bW_l-\bI\|+\|\ba_l\|$,
where $\bI$ is the identity matrix with the same dimensions as $\bW_l$. Accordingly, the overall network penalty is refined as   $\rho_3(\bvmu) = \rho_{31}(\bvmu)+\lambda_4 \|\btheta\|_1$.

As a result, four penalty parameters need to be tuned. Conducting a direct grid search over all parameters simultaneously can be computationally burdensome. Since variable selection is based on the extracted representations and the network complexity also should match the number of selected representations, we hence first determine the number of representations by tuning $\lambda_1$. Because we prefer networks with shallower structures, we then select $\lambda_3$ before tuning $\lambda_4$.
Consequently, we adopt a sequential grid search for $\lambda_1$, $\lambda_2$, $\lambda_3$, and $\lambda_4$ in a region around zero: (1) select $\lambda_1$ with $(\lambda_2, \lambda_3, \lambda_4)$ fixed at 0; (2) select $\lambda_2$ with $\lambda_1$ fixed at its chosen value and $(\lambda_3, \lambda_4)$ fixed at 0; (3) select $\lambda_3$ with $(\lambda_1,\lambda_2)$ fixed at their chosen values and $\lambda_4=0$; and (4) select $\lambda_4$ with $(\lambda_1,\lambda_2,\lambda_3)$ fixed at their chosen values.

\begin{figure}[ht]
    \centering
    \begin{minipage}[c]{0.33\textwidth}
        \begin{algorithm}[H]
            \spacingset{1}
            \footnotesize
            \normalem
            \KwIn{$\tau_1$, $\bB$}
            \KwOut{$\bB^{trunc}$ }
            Initialization $\bB^{trunc} \leftarrow \bB$\;
            \For{i=1:d}{
                \If{$\lVert \bB^{trunc}_{[i,]} \rVert_2 \le \tau_1 $}{
                    $ \bB^{trunc}_{[i,]} \leftarrow 0$\;
                }
            }
            Return $\bB^{trunc}$\;
            \caption{\footnotesize Parameter truncation algorithm}
            \label{alg:pta}
        \end{algorithm}
    \end{minipage}
    \hfill 
    \begin{minipage}[c]{0.66\textwidth}
        \begin{algorithm}[H]
        \spacingset{1}
            \footnotesize
            \normalem
            \KwIn{$\{\bX_i, Y_i\}_{i=1}^n$, $\lambda_1$, $\lambda_2$, $\lambda_3$ and $\lambda_4$}
            \KwOut{$\hat{\btheta}$, $\hat{\bB}$}
            Initialization $\btheta$, $\bB$\;
            \For{number of training iterations}{
                Sample mini-batch of $b$ samples $\{\bX_j, Y_j\}_{j=1}^b$ from dataset\;
                Compute the loss by \eqref{eq:group}\;
                Update $\btheta$ and $\bB$ by descending its stochastic gradient:
                $$
                \nabla_{\btheta, \bB} \ \frac{1}{b}\sum^b_{j=1} L(g_{\btheta}(\bB\bX_j), Y_j) + \blambda \brho(\bvmu),
                $$
                \If{truncating $\bB$ and $\btheta$}{
                    Truncating $\bB$ and $\btheta$ according to Algorithm \ref{alg:pta}\;
                }
            }
            Return $\hat{\btheta}$, $\hat{\bB}$\;
            \caption{\footnotesize Minibatch stochastic gradient descent training of the proposed method}
            \label{alg:all}
        \end{algorithm}
    \end{minipage}
\end{figure}



\section{Simulation study}
\label{sec:simulation}

We conduct simulation studies to evaluate the performance of the proposed method by comparing it with several state-of-the-art DNN-based approaches: the vanilla feed-forward neural network (DNN), deep representation selection with a selection layer (DFS; \citealp{chen2021nonlinear}), LassoNet \citep{lemhadri2021lassonet}, and the group concave regularization network (GCRNet; \citealp{luo2023sparse}). The vanilla DNN does not perform variable selection, whereas DFS, LassoNet, and GCRNet incorporate variable selection  within the DNN framework.
We further consider multi-index models with either more, fewer, or the same number of indices as the true model, where the link function is approximated by a DNN to alleviate the curse of dimensionality. 
A multi-index model with $k$ indices is refered to as index-model($k$). When $k$ equals the true number of indices, the corresponding model represents the oracle case, denoted as index-model($k^*$).
We adopt the ReQU activation function with $p=2$ for our method.

For regression problems, model performance is evaluated using
predictive error (PE) and mean square error (MSE) 
on the testing data,  defined as:
$\text{PE} = \frac{\sum^n_{i=1}{(\hat{Y}_i - Y_i)^2}}{\sum^n_{i=1} Y^2_i}$ and $\text{MSE} = \frac{1}{n} \sum^n_{i=1} (\hat{Y}_i - f_{0i})^2$, where $\hat{Y}_i$ and $Y_i$ denote the predicted and observed responses, repectively, $f_{0i}$ is the true regression function evaluated at the corresponding covariate, and $n$ is the test sample size.
For binary classification, performance is assessed using accuracy (ACC) and the area under the ROC curve (AUC):
$\text{ACC} = \frac{\sum^n_{i=1} \textbf{1}_{(\hat{Y}_i = Y_i)}}{n}$ and $\text{AUC}=\frac{\sum^{n_+}_{i=1} \sum^{n_-}_{j=1} \textbf{1}_{(p_i > p_j)}}{n_+ n_-}$, where $n_+$ and $n_-$ are the numbers of positive and negative samples, respectively, $p_i$ is the predicted probability of the positive class for the $i$th sample, and $\mathbf{1}_{(\cdot)}$ is the indicator function.
Model complexity is quantified by the extracted representation dimension (Dims.) and the proportion of zero elements (Prop.0).
Variable selection performance is evaluated using the true positive rate (TPR) and false positive rate (FPR), defined as $\text{TPR} = \frac{|\hat{S} \cap S^* |}{|S^*|}$ and $\text{FPR} = \frac{|\hat{S} \cap S^c |}{|S^c|}$,
where $\hat{S}$ represents the set of selected variables, $S^*$ the set of truly important variables, $S^c$ the set of truly unimportant variables, and $|\cdot|$ denotes set cardinality.
For each metric, we report the mean and standard deviation (SD) by repeating the experiment 50 times.

We consider four simulation settings in which data are generated from an additive regression model (Setting 1), an interactive regression model (Setting 2), an interactive regression model induced by a fully connected network (Setting 3), and an interactive classification model (Setting 4). Further details on the data generation process and experimental setup are provided in Supplementary F.1.

\subsection{Prediction and Selection Results}
The prediction and  selection results are summarized in Tables \ref{table:simulation} and Supplementary Tables G1, G2 and G3, which show that the proposed method has superior performance in prediction accuracy, representation dimension selection and variable selection across all four settings.
Specifically,

(1) The estimated dimension of the proposed method
is approximately 5 in Settings 1, 2, and 4, and around 10 in Setting 3, closely matching the true dimension. 
In contrast, except for multi-index models, the other competing methods do not incorporate dimension reduction, resulting in estimated dimensions nearly equal to the full number of variables.
On the other hand, the performance of the multi-index model is highly sensitive to the choice of the number of indices; both over-specification and under-specification can lead to significant performance degradation. Remarkably, our method is still superior to multi-index model with the oracle dimension. This is because our method not only reduces the dimension but also adaptively adjusts the network structure according to the selected dimension.
\begin{table}
\spacingset{0.95}
\centering
\caption{\footnotesize Simulation results for prediction, model complexity and variable selection for Setting 1. Predictive error (PE) and mean square error (MSE) are the model performance metrics. Dimension (Dims.) and proportion of zero elements in model parameters $\hat{\btheta}$ (Prop.0) are the metrics for model complexity. The true positive rate (TPR) and false positive rate (FPR) are the metrics for variable selection. All metrics are reported by the means with standard deviations in parentheses.}
\label{table:simulation}
\scalebox{0.6}{
\begin{tabular}{@{}ccccccccc@{}}
\toprule
             & $\rho$                & Methods & PE            & MSE            & Dims.           & Prop.0         & TPR            & FPR            \\ \midrule
\multirow{24}{*}{}  & \multirow{9}{*}{0}   & DNN     & 0.3559(0.0123) & 2.8494(0.0984) & 200             & 0.0392(0.0004) & 1              & 1              \\
                    &                      & Index model(1) & 0.3242(0.0037) & 2.6126(0.0315) &
                    1              & 0.0396(0.0004) &
                    1              &
                    1              \\
                    &                      & Index model(5$^*$) & 0.1665(0.0111) & 1.2747(0.0911) &
                    5              & 0.0390(0.0005) &
                    1              &
                    1              \\
                    &                      & Index model(10) & 0.1921(0.0124) & 1.4854(0.1043) &
                    10             & 0.0388(0.0004) &
                    1              &
                    1              \\
                    &                      & DFS & 0.3144(0.0517) & 2.4939(0.4316) & 102.8000(61.4505)  & 0.6972(0.0295) & 0.6770(0.2874) & 0.3510(0.3463) \\
                    &                      & LassoNet & 0.3578(0.0967) & 2.8710(0.8036) & 107.2600(28.5950)  & 0.1766(0.0016) & 0.8642(0.0957) & 0.2084(0.2004) \\
                    &                      & GCRNet  & 0.2752(0.0066) & 2.1778(0.0568) & 131.0500(16.6508)  & \textbf{0.7659(0.0170)} & 0.9295(0.0343) & 0.3810(0.1356) \\
                    &                      & DeepIn (Ours)  & \textbf{0.1557(0.0133)} & \textbf{1.1805(0.1126)} & \textbf{4.3600(0.8890)}  & 0.6586(0.2573) & \textbf{0.9556(0.0313)} & \textbf{0.0100(0.0100)} \\ \cmidrule(l){2-9}
                    & \multirow{9}{*}{0.1} & DNN     & 0.3656(0.0146) & 3.7708(0.1567) & 200             & 0.0388(0.0003) & 1              & 1              \\
                    &                      & Index model(1) & 0.3734(0.0928) & 3.8212(1.0132) &
                    1              & 0.0387(0.0006) &
                    1              &
                    1              \\
                    &                      & Index model(5$^*$) & 0.2293(0.0186) & 2.2886(0.2050) &
                    5              & 0.0386(0.0004) &
                    1              &
                    1              \\
                    &                      & Index model(10) & 0.2475(0.0241) & 2.4780(0.2593) &
                    10             & 0.0385(0.0003) &
                    1              &
                    1              \\
                    &                      & DFS & 0.4071(0.0535) & 4.1995(0.5655) & 120.4000(54.0355)  & 0.6781(0.0352) & 0.7498(0.2245) & 0.4542(0.3351) \\
                    &                      & LassoNet & 0.2939(0.0435) & 3.0041(0.4833) & 154.7800(18.7417)  & 0.1765(0.0009) & 0.9714(0.0462) & 0.5764(0.1470) \\
                    &                      & GCRNet  & 0.4286(0.0168) & 4.4476(0.1880) & 178.3500(23.6120)  & 0.6119(0.2470) & \textbf{0.9610(0.0512)} & 0.8225(0.1869) \\
                    &                      & DeepIn (Ours)  & \textbf{0.1724(0.0124)} & \textbf{1.6536(0.1328)} & \textbf{5.2800(0.9600)}  & \textbf{0.7789(0.1126)} & 0.9074(0.0207) & \textbf{0.0902(0.0463)} \\
                    \cmidrule(l){2-9}
                    & \multirow{9}{*}{0.2} & DNN     & 0.5282(0.0120) & 6.3029(0.1523) & 200             & 0.0392(0.0003) & 1              & 1              \\
                    &                      & Index model(1) & 0.4919(0.0221) & 5.8272(0.2804) &
                    1              & 0.0399(0.0004) &
                    1              &
                    1              \\
                    &                      & Index model(5$^*$) & 0.3029(0.0179) & 3.4996(0.2177) &
                    5              & 0.0390(0.0003) &
                    1              &
                    1              \\
                    &                      & Index model(10) & 0.3348(0.0197) & 3.8919(0.2425) &
                    10             & 0.0389(0.0003) &
                    1              &
                    1              \\
                    &                      & DFS & 0.4853(0.0356) & 5.7900(0.4460) & 111.2000(60.6841)  & 0.6880(0.0366) & 0.7340(0.2700) & 0.3780(0.3708) \\
                    &                      & LassoNet & 0.5413(0.0706) & 6.4769(0.8778) & 102.7600(26.6845)  & 0.1770(0.0005) & 0.8244(0.0920) & 0.2032(0.1804) \\
                    &                      & GCRNet  & 0.4571(0.0097) & 5.4194(0.1131) & 139.5500(13.0402)  & 0.7676(0.0135) & 0.9260(0.0206) & 0.4695(0.1145) \\
                    &                      & DeepIn (Ours)  & \textbf{0.2865(0.0146)} & \textbf{3.3098(0.1811)} & \textbf{4.2608(0.7922)}  & \textbf{0.8009(0.0002)} & \textbf{0.9590(0.0267)} & \textbf{0.0628(0.1927)} \\
                    \bottomrule
\end{tabular}
}
\end{table}

(2) While DFS and GCRNet exhibit good sparsity, our method attains substantially superior sparsity by jointly leveraging dimension reduction and adaptive network compression. This leads to a higher proportion of zero elements, thereby reducing statistical error and enhancing predictive accuracy. As shown in Table \ref{table:simulation} and Supplementary Tables G1 - G3 for Settings 1–4, substantial dimension reduction or architectural compression consistently yields notable gains in prediction performance.


(3) Regarding variable selection, our method achieves  the best performance  in removing noisy variables, as evidenced by its substantially lower FPR compared with DFS, LassoNet, and GCRNet. This effective noise reduction directly contributes to improved predictive accuracy across all settings. Moreover,  although LassoNet and GCRNet attain higher TPR than DFS in identifying informative variables, our method consistently outperforms all three competitors on this criterion as well.


In summary, the results in Table \ref{table:simulation} and Supplementary Tables G1, G2, and G3 highlight the strong dependence of predictive error on representation dimensionality, variable selection, and network architecture. By jointly optimizing these three components, our method achieves consistently superior performance across all four settings. In particular, relative to DNN, multi-index models, DFS, LassoNet, and GCRNet, it achieves average reductions in PE of 39.87\%, 42.84\%, and 88.16\%, and in MSE of 42.03\%, 46.23\%, and 91.24\% for regression tasks in Settings 1–3, respectively.
For the classification task in Setting 4, it improves ACC and AUC by  14.82\% and 11.45\% on average. 

\subsection{Hypothesis test}

\begin{figure}[ht]
\centering
    \subfigure[Setting 1 ($\rho=0.2$)]{
        \begin{minipage}[t]{0.48\linewidth}
        \centering
        \includegraphics[width=3in]{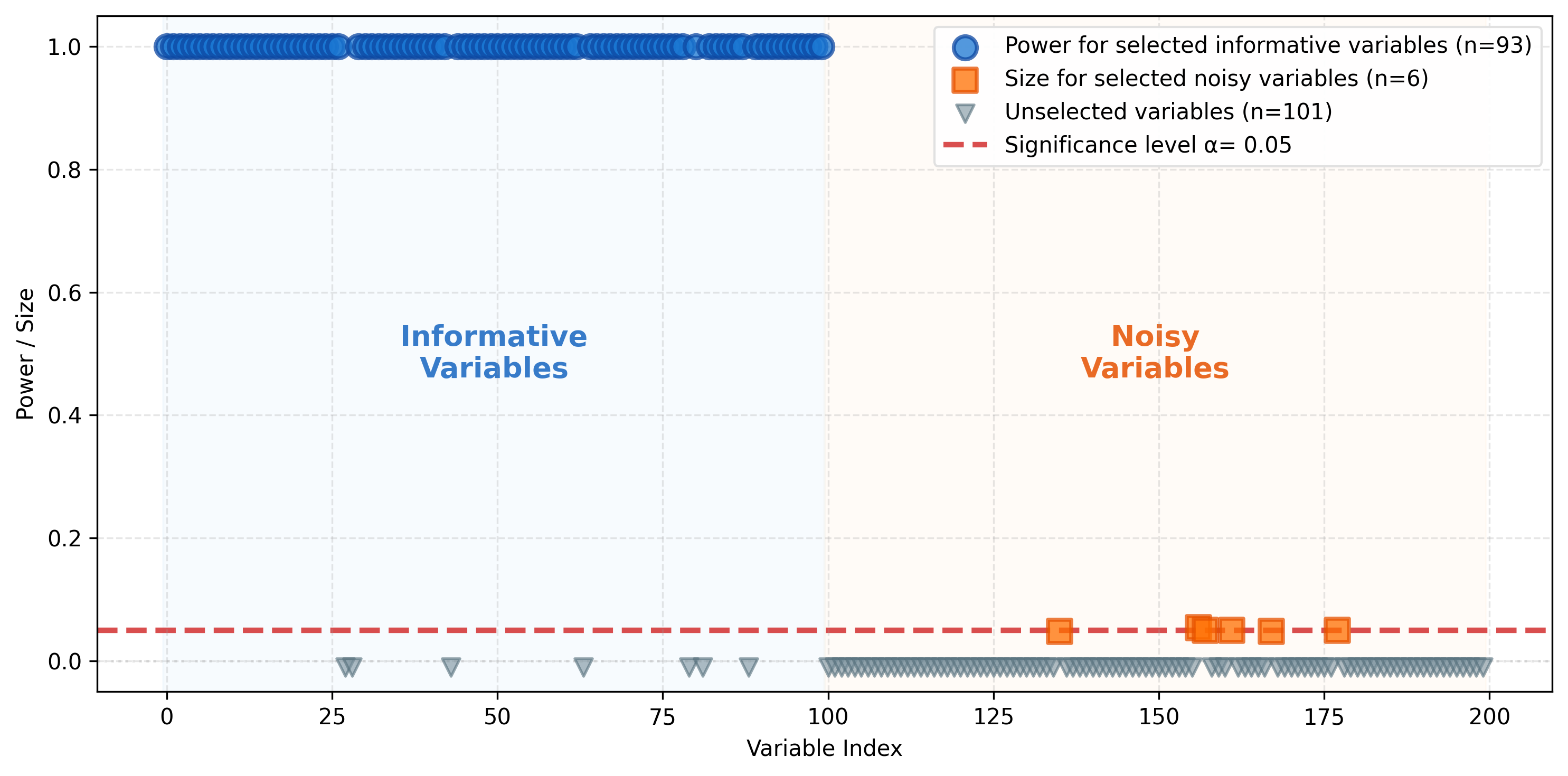}
        \end{minipage}%
    }
    \subfigure[Setting 2 ($\rho=0.2$)]{
        \begin{minipage}[t]{0.48\linewidth}
        \centering
        \includegraphics[width=3in]{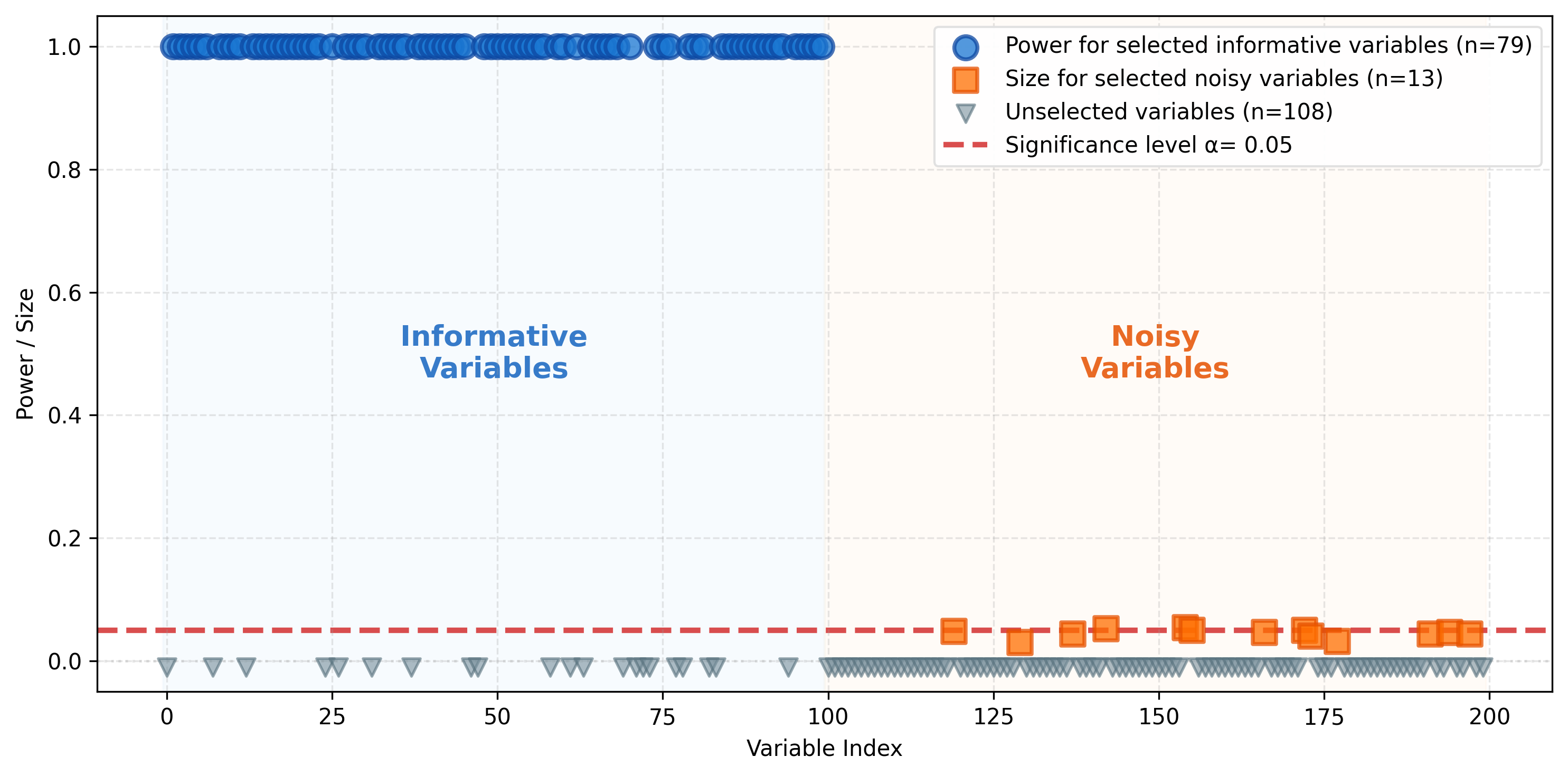}
        \end{minipage}%
    }

    \subfigure[Setting 3 ($\rho=0.2$)]{
        \begin{minipage}[t]{0.48\linewidth}
        \centering
        \includegraphics[width=3in]{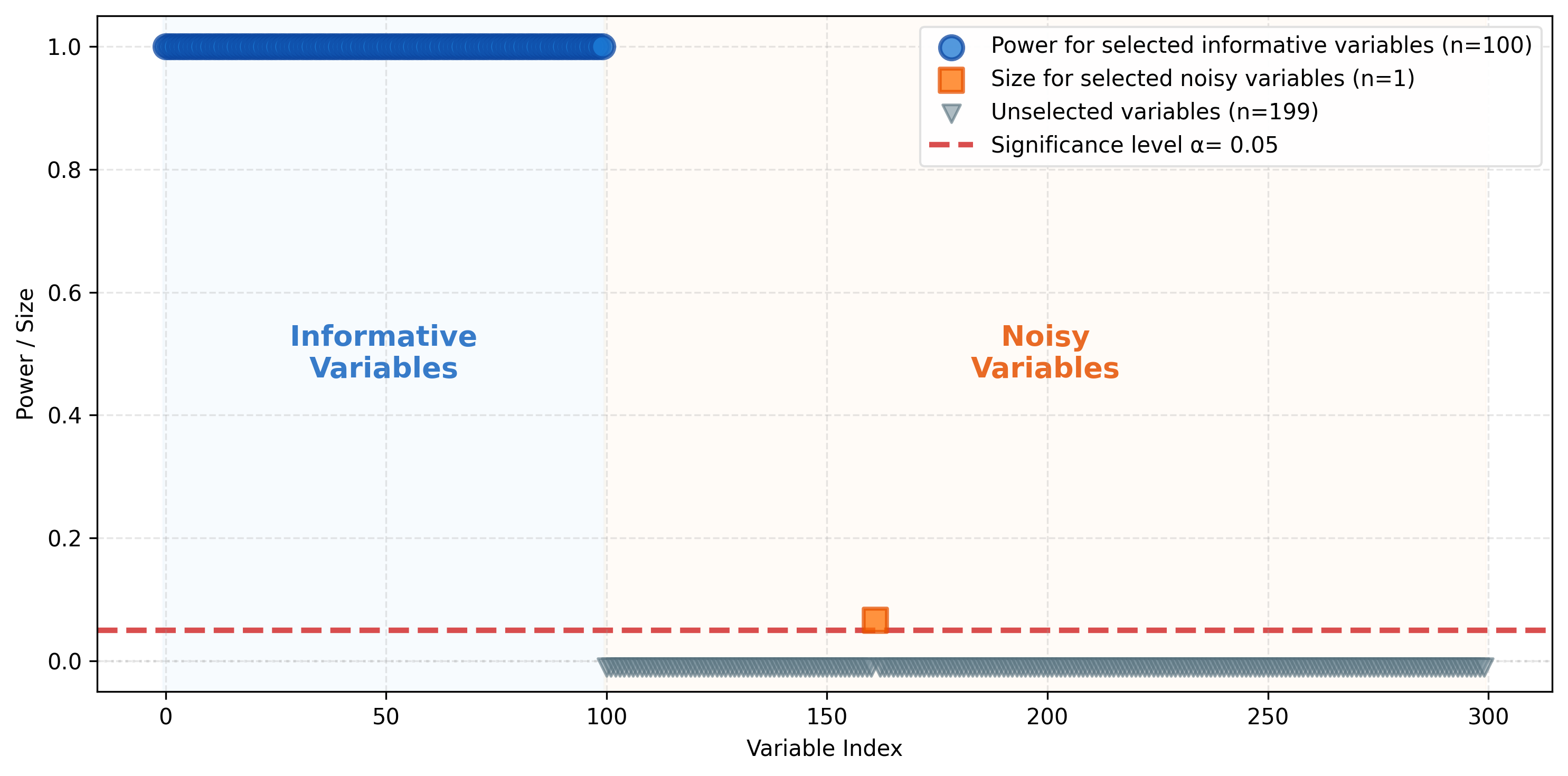}
        \end{minipage}%
    }
    \subfigure[Setting 4 ($\rho=0.2$)]{
        \begin{minipage}[t]{0.48\linewidth}
        \centering
        \includegraphics[width=3in]{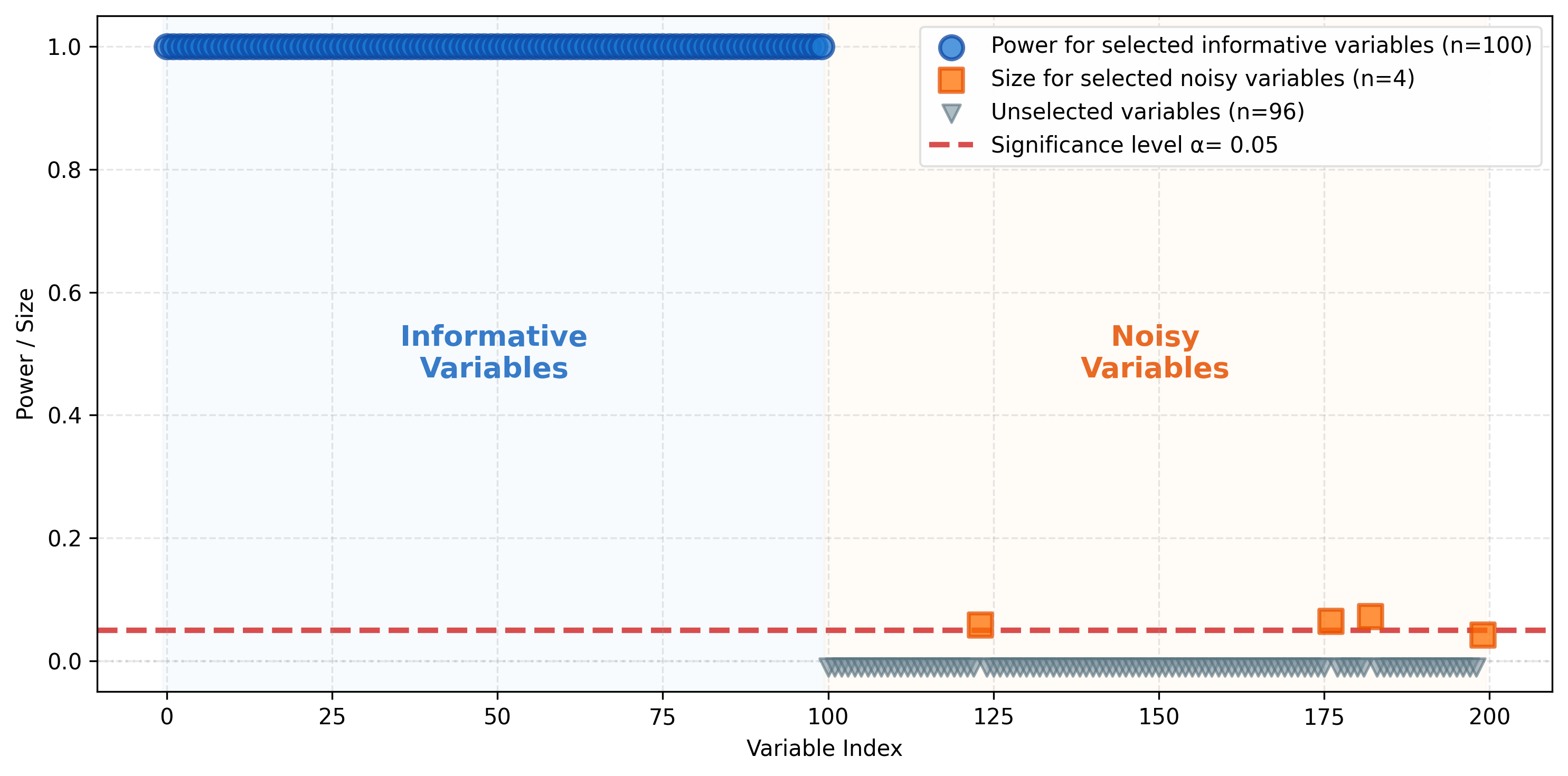}
        \end{minipage}%
    }
    \spacingset{0.95}
    \caption{{\footnotesize Test power for informative variables and empirical size for noisy variables in Settings 1–4 under correlation levels $\rho=0.2$.}
    }
    \label{fig:hypothesis}
\end{figure}

We first evaluate the performance of statistical inference for individual covariates.  Following variable selection, we compute the statistics $\{ \bU^{[j]} \}^{\hat{s}}_{j=1}$  500 times by bootstrap and Figure \ref{fig:hypothesis} reports the test power for the selected informative variables and the empirical size for the selected noisy variables under Settings 1-4 with $\rho=0.2$ at the significance level $\alpha=0.05$. The power for detecting informative variables consistently approaches 1 across all settings, confirming the method's high sensitivity to true effects. Meanwhile, the size for most noisy variables remains close to 0.05, demonstrating effective control of false positives and reflecting the consistency  of the test. Together, these results validate the proposed hypothesis test in accurately distinguishing informative variables from noise. Similar results for Settings 1-4 under correlation levels $\rho=0, 0.1$ are reported in Supplementary G.2.

\begin{figure}[htbp]
\centering
    \subfigure[]{
        \begin{minipage}[t]{0.3\linewidth}
        \centering
        \includegraphics[width=2in]{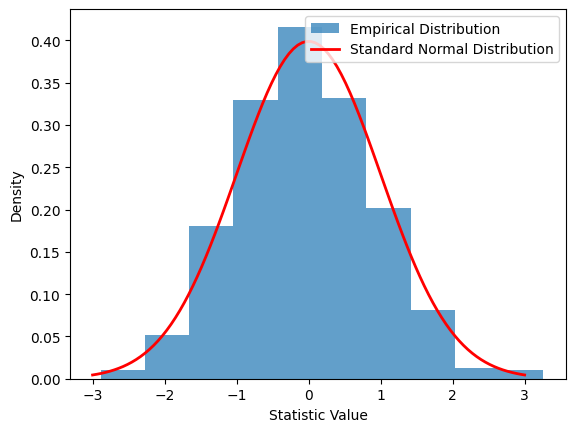}
        \end{minipage}%
    }
    \subfigure[]{
        \begin{minipage}[t]{0.3\linewidth}
        \centering
        \includegraphics[width=2in]{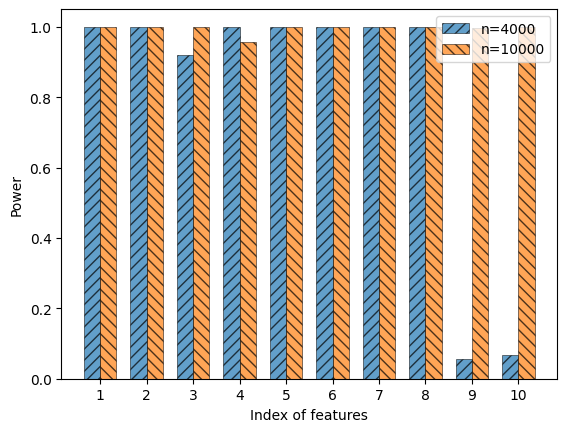}
        \end{minipage}%
    }
    \subfigure[]{
        \begin{minipage}[t]{0.3\linewidth}
        \centering
        \includegraphics[width=2in]{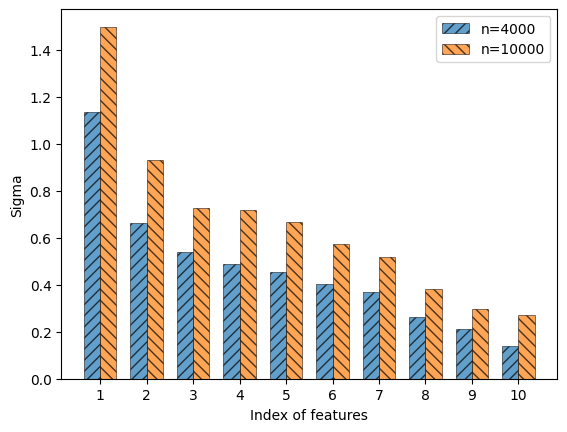}
        \end{minipage}%
    }
    \spacingset{0.95}
    \caption{{\footnotesize The distribution under $H_0$ and the power of different  representations with their singular value in Setting 3 ($\rho=0.2$): (a) The distribution under $H_0$; (b) The power of representations; (c) The singular value of representations.}
    }
    \label{fig:hypothesis_combine}
\end{figure}

Next, we evaluate the performance of statistical inference for representations, which are combinations of covariates. We first verify whether the proposed statistic follows a normal distribution under the null hypothesis, as established in Theorem \ref{thm:hypo}. Let $\mcI=\{1,\ldots,\hat{d}\}$ and compute the statistic $T_n/\hat\varsigma_2$ 1000 times under Setting 3 with $\rho=0.2$. Figure \ref{fig:hypothesis_combine}(a) shows that the empirical distribution of the statistic aligns closely with the standard normal distribution. Moreover, the empirical size of the test at the $\alpha=0.05$ significance level is 0.031, 
close to the nominal level.

We  test the significance of each learned representation. For each representation $j \in \{1,\ldots,\hat{d}\}$ with $\hat{d}=10$, we define $\mcI=\{1,\ldots,j-1,j+1,\ldots,\hat{d}\}$ and $\mcI^c={j}$. The resulting power under Setting 3 with $\rho=0.2$, based on 1000 replicates of 4000 samples, is shown by the blue bars in Figure \ref{fig:hypothesis_combine}(b). At the $\alpha=0.05$ level, the power exceeds 0.9 for the first eight representations, drops to about 0.1 for the ninth and tenth.
Figure \ref{fig:hypothesis_combine}(c) presents the corresponding normalized singular values (blue bars). The low power of the last two representations is consistent with their small singular values, which is expected since removing representations with negligible singular values has little effect on the resulting function. Increasing the sample size enhances their signals: as shown by the orange bars in Figures \ref{fig:hypothesis_combine}(b) and \ref{fig:hypothesis_combine}(c), raising the sample size to 10000 substantially improves both the power and singular values of these representations. Similar results on the empirical distribution   and the power of the statistic under other settings are reported in Supplementary  G.2.

\section{Real data example}
\label{sec:real_data}

In this section, we evaluate the performance of the proposed method on four real datasets: ADNI, BlogFeedback, MNIST, and FashionMNIST. 
During implementation, we adopt the same network architecture across all methods in each dataset.
For the proposed method,  the hyper-parameters $\blambda$ are tuned by the procedure described in Section~\ref{sec:imple} and are selected by minimizing MSE or ACC on the validation set.
The detailed settings of all hyper-parameters are provided in Supplementary F.2.
For each metric, we report the mean and standard deviation (SD) by repeating the experiment 10 times.


\subsection{ADNI dataset}


Alzheimer's disease (AD) is a prevalent and irreversible neurodegenerative disorder in the elderly, characterized by progressive deficits in memory, cognition, language, and reasoning. The Alzheimer's Disease Neuroimaging Initiative (ADNI) is a longitudinal, multicenter study designed to develop and validate biomarkers for the early diagnosis and progression monitoring of AD.
Following standard quality control procedures, our analysis includes 295 samples from ADNI-GO and ADNI2, two phases of the ADNI study. ADNI-GO (2010–2011) focused on early mild cognitive impairment and amyloid PET imaging, whereas ADNI2 (2011–2016) expanded the cohort and incorporated tau PET imaging to enable more comprehensive biomarker evaluation.
Our goal is to assess whether genetic factors can predict the progression of mild cognitive impairment and early-stage AD. After removing transcripts that could not be annotated, we retained 48,157 gene expression profiles from an initial set of 49,293 transcripts. The Mini-Mental State Examination (MMSE), ranging from 0 to 30, serves as the response variable, with lower scores indicating greater cognitive impairment. Following \cite{chen2021nonlinear}, we further applied marginal representation screening to reduce the number of genes to 192.

The top block of Table~\ref{table:result_reg} summarizes the comparative results of the seven methods for ADNI study. The proposed method
outperforms all competitors in prediction accuracy, as indicated by lower PE and MSE, and attains these gains using substantially fewer representations and variables.
Notably, the multi-index model with an appropriate number of indices still underperforms the DNN. This is likely because the limited sample size in the ADNI study is insufficient to learn the multi-index model. In contrast, our method integrates dimensionality learning, variable selection, and adaptive network compression, which together substantially reduce the total number of parameters and consequently the statistical error, while controlling the approximation error, thereby achieving superior PE and MSE performance  even with a small sample size.

We selected 64 genes based on the intersection of repeated experiments, with their corresponding p-values reported in Table~\ref{table:gene_pvalue}. Notably, several of these genes are known to potentially have an impact on AD pathogenesis. For example, mutations in GBA impair lysosomal function, reducing the clearance of $\alpha$-synuclein and $\beta$-amyloid  aggregates, thereby possible promoting neurotoxicity and accelerating AD progression  \citep{ huttenrauch2018glycoprotein,mata2016gba}.
EIF2AK3 (PERK) mediates the unfolded protein response during endoplasmic reticulum stress, leading to sustained eIF2$\alpha$ phosphorylation that suppresses protein translation while increasing BACE1 expression, thereby promoting $\beta$-amyloid production and tau pathology \citep{wong2019eif2ak3}. Other genes, including ARSA, SCN1B, STIM1, and ALDOA, may influence AD through pathways such as myelination, neuronal excitability, and metabolic dysfunction \citep{makarious2019arsa, williams2024development, tinsley2024kmt2c}.
These genes are highlighted in Table \ref{table:gene_pvalue}, with all showing statistically significant associations ($p < 0.05$), providing strong evidence for their involvement in AD pathogenesis.

\begin{table}[thbp]
\centering
\spacingset{0.95}
\caption{\small Variable selection and prediction performance for the ADNI and BlogFeedback datasets. ``\#Variables" indicates the number of selected variables. All metrics are reported by the means with standard deviations in parentheses.}
\label{table:result_reg}
\scalebox{0.68}{
\begin{tabular}{@{}ccccccc@{}}
\toprule
Dataset               & Methods  & PE            & MSE            & Dims.           & \#Variables       & Prop.0         \\ \midrule
\multirow{7}{*}{ADNI} & DNN      & 0.4314(0.1102) & 5.5006(1.4050) & 192  & 192      & 0.0390(0.0009) \\
                      & Index model(10) & 0.7793(0.1749) & 9.9376(2.2303) & 10
                      & 192 & 0.0407(0.0023) \\
                      & Index model(100) & 0.5390(0.1006) & 6.8727(1.2824) & 100
                      & 192 & 0.0341(0.0004) \\
                      & Index model(150) & 0.5580(0.0832) & 7.1163(1.0608) & 150
                      & 192 & 0.0297(0.0008) \\
                      & DFS & 0.8108(0.0586) & 10.3398(0.7477) & 170.6667(12.2701)
                      & 170.6667(12.2701) & \textbf{0.4461(0.0350)} \\
                      & LassoNet & 0.6099(0.1651) & 7.7778(2.1055) & 163.4444(32.7960)
                      & 163.4444(32.7960) & 0.1719(0.0733) \\
                      & GCRNet   & 0.4456(0.0552) & 5.6829(0.7041) & 192
                      & 192 & 0.3711(0.1384) \\
                      & DeepIn (Ours)     & \textbf{0.4164(0.0588)} & \textbf{5.3105(0.7495)} & \textbf{137.6000(46.1589)}
                      & \textbf{138.7000(44.4906)}  & 0.1045(0.0222)             \\ \hline
\multirow{7}{*}{Blog} & DNN      & 0.4239(0.0107) & 0.5368(0.0135) & 280  & 280           & 0.0386(0.0011) \\
                      & Index model(10) & 0.4009(0.0025) & 0.5077(0.0031) & 10
                      & 280 & 0.0386(0.0004) \\
                      & Index model(100) & 0.4150(0.0070) & 0.5255(0.0089) & 100
                      & 280 & 0.0376(0.0011) \\
                      & Index model(150) & 0.4171(0.0075) & 0.5282(0.0095) & 150
                      & 280 & 0.0375(0.0007) \\
                      & DFS & 0.5652(0.0236) & 0.7157(0.0299) & 137.2000(86.2564) & 137.2000(86.2564) & 0.4738(0.0133) \\
                      & LassoNet & 0.4924(0.0232) & 0.6236(0.0293) & 32.6000(10.7350) & 32.6000(10.7350) & 0.2259(0.0009) \\
                      & GCRNet & 0.4615(0.0046) & 0.5845(0.0058) & 180.1000(84.8121) & 180.1000(84.8121) & 0.5002(0.1035) \\
                      & DeepIn (Ours)     & \textbf{0.3633(0.0039)} & \textbf{0.4601(0.0050)} & \textbf{11.0000(1.7889)} & \textbf{30.8000(4.0939)} & \textbf{0.7359(0.0189)} \\ \bottomrule
\end{tabular}
}
\end{table}

\begin{table}[htbp]
\centering
\caption{The selected genes and their p-values for ADNI analysis.}
\label{table:gene_pvalue}
\scalebox{0.64}{
\begin{threeparttable}
\begin{tabular}{cc|cc|cc|cc|cc|cc} 
\toprule
Gene  &p-value & Gene  &p-value & Gene  &p-value & Gene  &p-value & Gene  &p-value & Gene  &p-value                       \\ \midrule
RAB3D & 0.0000 & NOP14 & 0.0000 & TBX5 & 0.0000 & KMT2C & 0.087 & XPO1 & 1.0000 & RAB1B & 0.5291 \\
\textbf{STIM1} & \textbf{0.0000} & LANCL1 & 0.0043 & CLPTM1 & 0.0000 & FMO1 & 0.0029 & \textbf{ARSA} & \textbf{0.0000} & DCAF8 & 0.0000 \\
MICU1 & 0.3492 & MTSS1 & 0.9971 & MAPKAP1 & 0.0000 & MAOB & 0.0245 & TSC22D4 & 0.0000 & LOC $\|$ RAB \tnote{a} & 0.0000 \\
TBPL1 & 0.0000 & DIO2 & 0.0384 & \textbf{EIF2AK3} & \textbf{0.0000} & \textbf{GBA} & \textbf{0.0048} & CHAC1 & 0.0000 & AKAP11 & 0.4741 \\
PRR13 & 0.0000 & BTAF1 & 1.0000 & \textbf{SCN1B} & \textbf{0.0000} & NAPEPLD & 0.0000 & ATP6AP1 & 0.0000 & CLCN1 & 0.0000 \\
TMEM127 & 0.0021 & IFT74 & 0.0058 & HAPLN2 & 0.0000 & ACADM & 0.0003 & PLCH1 & 0.0000 & EPS15L1 & 0.5907 \\
PPP5C & 0.0563 & TMED2 & 0.0012 & FANCF & 0.0018 & ALAS1 & 0.9997 & ATG13 & 0.0000 & FAM160A1 & 0.0000 \\
LIG3 & 0.0000 & TMEM110 & 0.0000 & PDLIM7 & 0.0682 & TIRAP & 0.0029 & NRL & 0.0029 & FOXD2 & 0.0000 \\
ARNT & 0.4124 & OR1C1 & 0.0000 & FLOT1 & 0.0000 & C15ORF48 & 0.0000 & CEP57 & 0.0000 & FAM69A & 0.0316 \\
TBCD & 0.0000 & TRABD & 0.0000 & AIDA & 0.0146 & NKRF & 0.5566 & MYT1L & 0.0000 & LILRB3 & 1.0000 \\
RDM1 & 0.0000 & SLC39A1 & 0.0000 & PACSIN2 & 0.0000 & \textbf{ALDOA} & \textbf{0.0000} & & & & \\
\bottomrule
\end{tabular}
\begin{tablenotes}    
\footnotesize               
\item[a] This is the abbreviation of gene LOC101929253 $\|$ RAB40C.          
\end{tablenotes}            
\end{threeparttable}
}
\end{table}

\subsection{BlogFeedback dataset}

The BlogFeedback dataset, obtained from the UCI Machine Learning Repository,
contains feature representations derived from blog posts for predicting the number of comments each post receives. The response variable records the number of comments generated within 24 hours after publication. The dataset includes 280 features (detailed in Supplementary F.3) and 52,397 samples in total.

As shown in the bottom block of Table \ref{table:result_reg}, our method
outperforms all competing approaches. In particular,  it achieves the highest prediction accuracy while using the fewest dimensions and variables as well as the sparsest network, again highlighting its capacity to uncover a low-dimensional latent space based on the selected appropriate variables and  network. 

To assess the stability of variable selection,
we compare the frequencies of the top-selected variables across 10 repeated experiments. Supplementary Table G4 
reports the 20 most frequently selected variables and their frequencies. Our method shows highest consistency in variable selection across repetitions.
In addition, we test the significance of the learned  representations according to Theorem \ref{thm:hypo}. As shown in Supplementary Figure G4, the empirical power based on 100 replicates is close to 1 for each representation, indicating that all learned representations play an important role in predicting the response.

\subsection{MNIST and FashionMNIST dataset}

Both the MNIST and FashionMNIST datasets 
are 10-class classification problems.
The MNIST dataset consists of grayscale images of handwritten digits (0–9). Each image is represented as a $28\times 28$ pixel matrix, with pixel intensities normalized to values between 0 and 1. The FashionMNIST dataset, a more challenging alternative to MNIST, includes grayscale images representing ten distinct clothing and accessory classes, such as T-shirts, dresses, shoes, and bags.
Both the MNIST and FashionMNIST training datasets  contain 60,000 images while the testing datasets contain 10,000 images. We split the training dataset into 55,000 and 5,000 for training and validating sets.

To satisfy the input requirement of the framework of DNNs, we flatten each $28 \times 28$ image pixel matrix into 784-dimensional vector. Since GCRNet is designed for binary classification and the multi-index model mainly for regression, we compare our method with vanilla DNN, DFS and LassoNet.


\begin{table}[htbp]
\spacingset{0.95}
\centering
\caption{\small The variable selection and prediction performance of MNIST and Fashion MNIST dataset. ``\#Variables" means the number of selected variables. All metrics are reported by the means with standard deviations in parentheses.}
\label{table:result_mnist}
\scalebox{0.7}{
\begin{tabular}{@{}cccccc@{}}
\toprule
Dataset                       & Methods  & Acc.           & Dims.    & \#Variables         & Prop.0         \\ \midrule
\multirow{4}{*}{MNIST}        & DNN      & 0.9583(0.0045) & 784    & 784           & 0.0360(0.0003) \\
                              & DFS & 0.9526(0.0101) & 430.8000(169.0330) & 430.8000(169.0330) & 0.5550(0.0226) \\
                              & LassoNet & 0.9508(0.0092) & 413.2000(171.1723) & \textbf{413.2000(171.1723)} & 0.2818(0.1219) \\
                              & DeepIn (Ours)     & \textbf{0.9730(0.0011)} & \textbf{331.6000(62.6086)} & 457.4000(3.7736) & \textbf{0.5916(0.0092)} \\ \hline
\multirow{4}{*}{FashionMNIST} & DNN      & 0.8874(0.0032) & 784      & 784         & 0.0329(0.0007) \\
                              & DFS & 0.8786(0.0071) & 470.0000(240.4483) & \textbf{470.0000(240.4483)} & \textbf{0.5022(0.0411)} \\
                              & LassoNet & 0.8408(0.0168) & 540.7000(128.7929) & 540.7000(128.7929) & 0.2358(0.0904) \\
                              & DeepIn (Ours)     & \textbf{0.8920(0.0016)} & \textbf{460.6000(85.2282)} & 716.4000(4.5431) & 0.3172(0.1360) \\ \bottomrule
\end{tabular}
}
\end{table}

As shown in Table \ref{table:result_mnist}, our method achieves the highest prediction accuracy while using the fewest representations. Notably, both DFS and LassoNet perform worse than a vanilla DNN. This may be because DFS and LassoNet rely heavily on sparsity assumptions.  For image data, however, individual pixels carry little information, and meaningful features typically arise from combinations of pixels (e.g., lines, shapes, textures). Under a strict sparsity assumption, such combined effects tend to be ignored, leading to suboptimal performance. In contrast, our method explicitly captures these joint effects through the matrix $\bB$. Although this results in selecting more variables, the learned representations are more informative, yielding superior predictive accuracy.

\begin{figure}[htbp]
\centering
\begin{minipage}[t]{0.48\textwidth}
    \centering
    \begin{minipage}[t]{0.32\linewidth}
        \centering
        \includegraphics[width=0.87in]{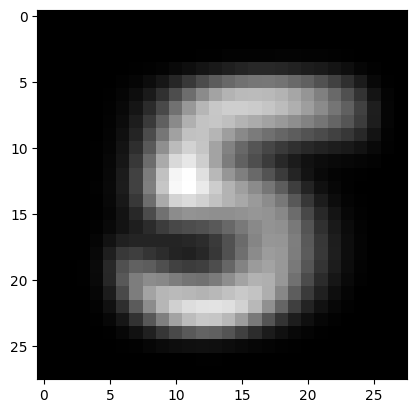}
    \end{minipage}%
    \begin{minipage}[t]{0.32\linewidth}
        \centering
        \includegraphics[width=1in]{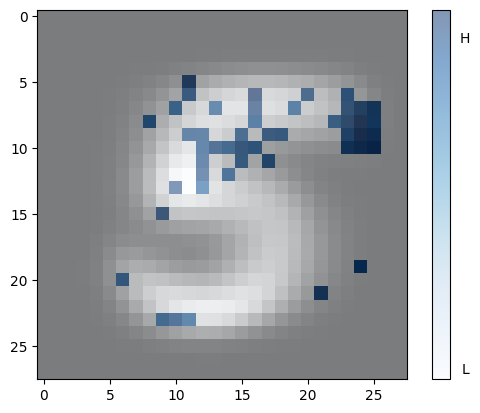}
    \end{minipage}%
    \begin{minipage}[t]{0.32\linewidth}
        \centering
        \includegraphics[width=1in]{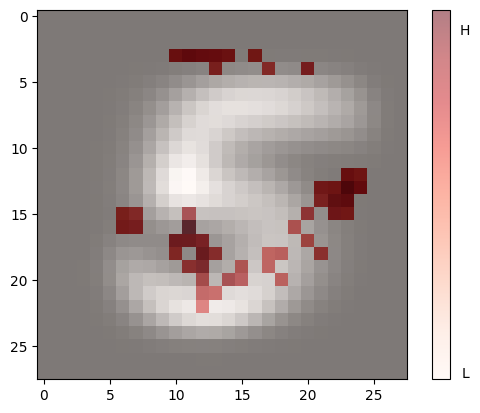}
    \end{minipage}%

    \vspace{0.5em} 

    \begin{minipage}[t]{0.32\linewidth}
        \centering
        \includegraphics[width=0.87in]{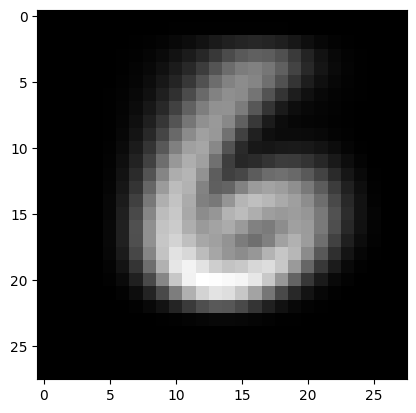}
    \end{minipage}%
    \begin{minipage}[t]{0.32\linewidth}
        \centering
        \includegraphics[width=1in]{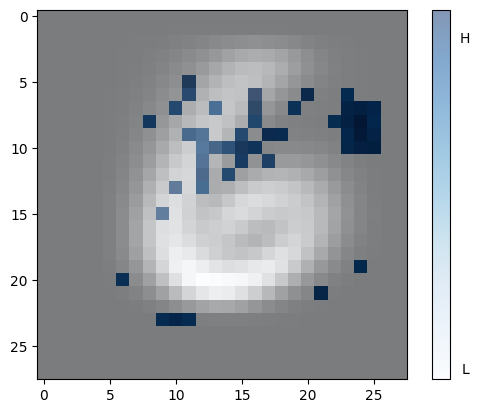}
    \end{minipage}%
    \begin{minipage}[t]{0.32\linewidth}
        \centering
        \includegraphics[width=1in]{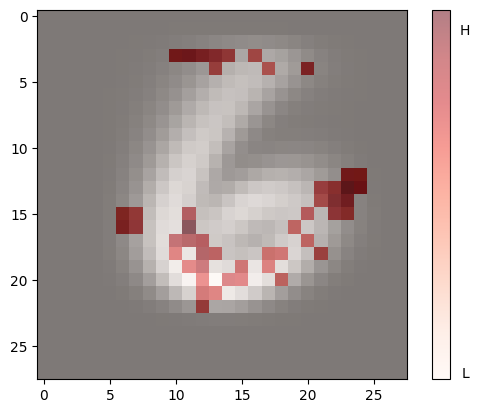}
    \end{minipage}%
    \spacingset{0.95}
    \caption{\footnotesize \textbf{First column:} The digit 5 and 6.
    \textbf{Second column:} Important pixels for predicting digit 5.
    \textbf{Last column:} Important pixels for predicting digit 6.
    }
    \label{fig:variable_selection_mnist}
\end{minipage}%
\hfill
\begin{minipage}[t]{0.48\textwidth}
    \centering
    \begin{minipage}[t]{0.32\linewidth}
        \centering
        \includegraphics[width=0.87in]{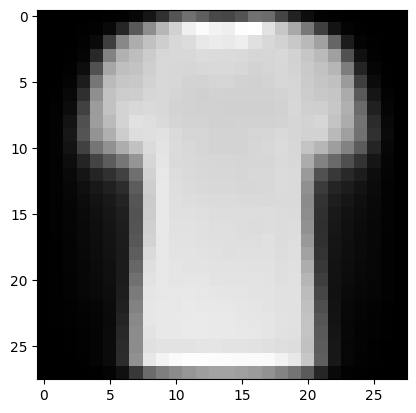}
    \end{minipage}%
    \begin{minipage}[t]{0.32\linewidth}
        \centering
        \includegraphics[width=1in]{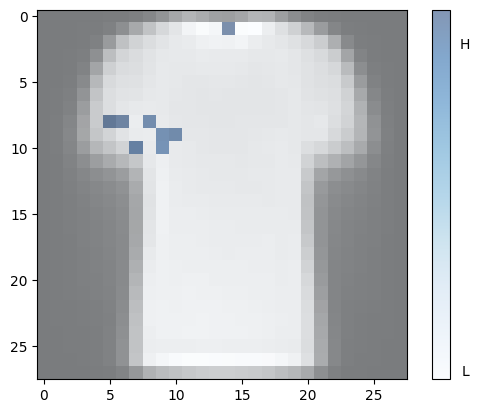}
    \end{minipage}%
    \begin{minipage}[t]{0.32\linewidth}
        \centering
        \includegraphics[width=1in]{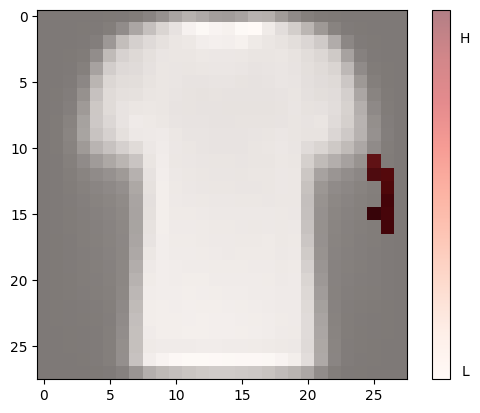}
    \end{minipage}%

    \vspace{0.5em} 

    \begin{minipage}[t]{0.32\linewidth}
        \centering
        \includegraphics[width=0.87in]{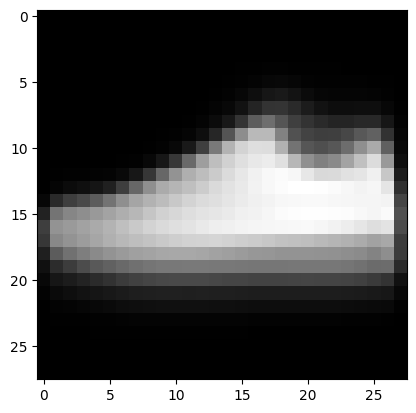}
    \end{minipage}%
    \begin{minipage}[t]{0.32\linewidth}
        \centering
        \includegraphics[width=1in]{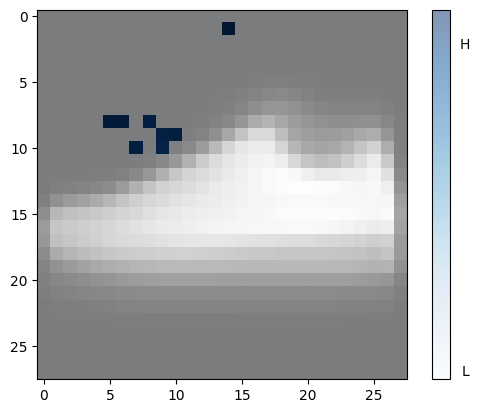}
    \end{minipage}%
    \begin{minipage}[t]{0.32\linewidth}
        \centering
        \includegraphics[width=1in]{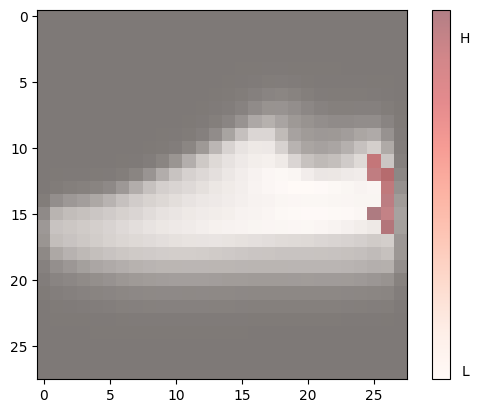}
    \end{minipage}%
    \spacingset{0.95}
    \caption{\footnotesize \textbf{First column:} The image T-shirt and sneaker.
    \textbf{Second column:} Important pixels for predicting T-shirt.
    \textbf{Last column:} Important pixels for predicting sneaker.
    }
    \label{fig:variable_selection_fashion_mnist}
\end{minipage}
\end{figure}

To further assess our method's variable selection capability, we use two binary subsets: digits 5 and 6 from MNIST, and T-shirts and sneakers from FashionMNIST.
Figure \ref{fig:variable_selection_mnist} illustrates variable selection results  for digits 5 and 6. The first column presents the average images of these digits from the training set, while the second and third columns display the discriminative pixels identified by our method for distinguishing digits 5 and 6, respectively. Additional details regarding the implementation are provided in Supplementary F.4.
From Figure \ref{fig:variable_selection_mnist}, we observe 
that, for digit 5, the model primarily focuses on the head region at the upper right, the neck area immediately above, and the horizontal stroke at the bottom. 
All of these areas are blank regions in digit 6. Conversely, the classification of digit 6 relies on the connected region in the lower-left quadrant and the right-side curve, both of which are inactive in digit 5.

This discriminative capability is also evident in FashionMNIST, as shown in Figure \ref{fig:variable_selection_fashion_mnist}. For T-shirt identification, our method focuses on the sleeve and collar regions, which are absent in sneaker images. In contrast, sneaker classification emphasizes the heel region, which is inactive in T-shirt images. These mutually exclusive activation patterns across both datasets consistently demonstrate the effectiveness of our method in identifying class-specific discriminative representations for image classification.

\section{Conclusion}
\label{sec:conclusion}

In this study, we introduce a novel neural network framework, DeepIn, for high-dimensional learning. The key innovation is the incorporation of a matrix $\bB$ into vanilla DNNs. By imposing group penalties on the columns and rows of $\bB$, the framework simultaneously performs dimension reduction, representation learning, and variable selection. In addition, a tailored architecture penalty compresses the network according to the learned representation dimension, reducing statistical error and enhancing predictive performance.
Rigorous theoretical analysis establishes the consistency of DeepIn in terms of representation dimension, variable selection, and network architecture. We further derive optimal non-asymptotic error bounds for the DeepIn estimator of order $\hat d^{\beta \vee (1+\lfloor\beta\rfloor)}(n/\log^c n)^{-2\beta/(\hat d+2\beta)}$, where $\hat d$ is the number of learned representations. This result highlights the critical importance of selecting low-dimensional representations. Based on this selection consistency, we also develop a novel hypothesis testing procedure to evaluate the significance of selected variables and learned representations.
Extensive numerical experiments on four benchmark datasets demonstrate the superior performance of DeepIn in dimension reduction, variable selection, network identification, and statistical inference, resulting in substantial gains in both predictive accuracy and interpretability. Notably, the method automatically uncovers human-interpretable discriminative patterns, such as characteristic strokes in digit recognition or garment-specific regions in fashion item classification.


A valuable extension of our method is domain-specific variable selection, such as group-level selection, which is particularly important in fields like biomedicine and social sciences, where prior knowledge often defines meaningful variable sets \citep{huang2012selective, buch2023systematic}. A natural way to incorporate such knowledge is by grouping variables and applying penalties to these groups of columns rather than to individual columns. Our framework can seamlessly accommodate such scenarios to achieve group-level variable selection. Additionally, incorporating a simple sparsity-inducing penalty on each element of $\bB$ could further reduce the number of parameters in ultra-high-dimensional settings and help identify the meaning of each learned representation. 

\section*{Data availability}
The datasets used in this study are publicly accessible. Specifically, ADNI dataset can be found in \url{http://www.adni-info.org/}, the BlogFeedback dataset can be obtained from the UCI Machine Learning Repository at \url{https://archive.ics.uci.edu/}, and MNIST and FashionMNIST datasets can be directly downloaded from PyTorch.

\section*{Acknowledgements}
The research was supported by National Key R\&D Program of China (No.2022YFA1003702), National Natural Science Foundation of China (Nos.12426309, 12171374, 12371275), Sichuan Science and Technology Program, China (Grant No. 2025JDDJ0007), Opening Project Fund of National Facility for Translational Medicine (Shanghai), New Cornerstone Science Foundation, and Guanghua Talent Project of SWUFE.

\normalem
\bibliography{ref,reference}

\end{document}